\documentclass[a4paper,fleqn,usenatbib]{mnras}

\usepackage{newtxtext,newtxmath}
\usepackage{enumitem}

\usepackage[T1]{fontenc}
\usepackage{ae,aecompl}

\usepackage[dvipdfmx]{graphicx}	
\usepackage{amsmath}	
\usepackage{amssymb}	

\title[A New Possible Accretion Scenario for ULXs]{A New Possible Accretion Scenario for Ultra-Luminous X-ray Sources}

\author[S. B. Kobayashi et al.]{
Shogo B. Kobayashi,$^{1}$\thanks{E-mail: shogo.kobayashi@rs.tus.ac.jp}
K. Nakazawa,$^{2}$
and K. Makishima$^{3,4,5}$
\\
$^{1}$Department of Physics, Tokyo University of Science, 1-3 Kagurazaka, Shinjuku-ku, Tokyo, 162-8601, Japan\\
$^{2}$Division of Particle and Astrophysical Science, and Kobayashi-Maskawa Institute for the Origin of Particles and the Universe,\\ Nagoya University, Furo-cho Chikusa-ku, Nagoya, 464-8602, Japan\\
$^{3}$High Energy Astrophysics Laboratory, The Institute of Physics and Chemical Research (RIKEN), 2-1 Hirosawa, Wako, Saitama,\\ 351-0198, Japan\\
$^{4}$Department of Physics, The University of Tokyo, 7-3-1 Hongo, Bunkyo-ku, Tokyo, 113-0033, Japan\\
$^{5}$Kavli Institute for the Physics and Mathematics of the Universe, The University of Tokyo, 5-1-5 Kashiwa-no-ha, Kashiwa, Chiba, 277-8683,\\ Japan 
}

\date{Accepted XXX. Received YYY; in original form ZZZ}

\pubyear{2017}

\begin{document}
\label{firstpage}
\pagerange{\pageref{firstpage}--\pageref{lastpage}}
\maketitle

\begin{abstract}
Using archival data from \textit{Suzaku}, \textit{XMM-Newton}, and \textit{NuSTAR}, nine representative Ultra-Luminous X-ray sources (ULXs) in nearby galaxies were studied. Their X-ray spectra were all reproduced with a multi-color disk emission model plus its Comptonization. However, the spectral shapes of individual sources changed systematically depending on the luminosity, and defined three typical spectral states. These states differ either in the ratio between the Comptonizing electron temperature and the innermost disk temperature, or in the product of Compton y-parameter and fraction of the Comptonized disk photons. The luminosity range at which a particular state emerges was found to scatter by a factor of up to 16 among the eight ULXs. By further assuming that the spectral state is uniquely determined by the Eddington ratio, the sample ULXs are inferred to exhibit a similar scatter in their masses. This gives a model-independent support to the interpretation of ULXs in terms of relatively massive black holes. None of the spectra showed noticeable local structures. Especially, no Fe K-shell absorption/emission lines were detected, with upper limits of $30\--40$ eV in equivalent width from the brightest three among the sample; NGC 1313 X-1, Holmberg IX X-1, and IC 342 X-1. These properties disfavor ordinary mass accretion from a massive companion star, and suggest direct Bondi-Hoyle accretion from dense parts of the interstellar medium.
\end{abstract}




\section{Introduction}

By the end of 2017 August, the gravitational wave telescopes LIGO and Virgo have so far  detected 10 gravitational-wave events from merging black holes (BHs) \citep{Abbott2016a, Abbott2016b, Abbott2017a, Abbott2017b, LIGO2018}. Out of the 20 component BHs involved in these events, 15 have masses higher than 20 $M_{\rm \odot}$, where $M_{\rm \odot}$ is the solar mass. Furthermore, the 20 remnant BHs have a rather flat mass distribution from $18  M_{\rm \odot}$  to $80 M_{\rm \odot}$. Clearly, the mass range of these BHs is significantly higher than that of BHs in X-ray emitting BH binaries (BHBs), namely, $5\--16 \ M_{\rm{\odot}}$ \citep{Ozel2010}, and that of theoretically predicted BHs (e.g., \citealt{Belczynski2010}) as an endpoint of massive stars (except Population III ones or those in a low metallicity environment). 
These newly recognized objects can potentially fill the gap in the BH mass distribution between $\sim20 M_{\rm{\odot}}$ to $\sim10^{4}M_{\rm{\odot}}$. We are hence encouraged to search other astrophysical phenomena for similar ``massive stellar BHs'', or intermediate-mass BHs (IMBHs).

One candidate for such BHs is extra-Galactic X-ray emitters called Ultra-Luminous X-ray Sources (ULXs), as initially suggested by \citet{Makishima2000}.
These are located on the arms of spiral galaxies \citep{Fabbiano1987}, and their luminosities, $L_{\rm{X}}=10^{39\--41}$~erg~sec$^{-1}$, often exceed significantly the Eddington luminosity $L_{\rm{edd}}$ of stellar mass ($\sim10 \ M_{\rm{\odot}}$) BHs. 
Supposing that ULXs are emitting X-rays in the sub-Eddington regime, they are expected to harbor the missing intermediate-mass BHs as their central objects \citep{Makishima2000}. 
However, their X-ray spectra appear different from what are observed from Galactic/Magellanic BH binaries. 
This fact, together with yet unknown scenario of the formation of BHs in such a mass range, led many authors to consider alternative interpretations, such as emission from stellar mass BHs that are accreting at super-critical rates (e.g., \citealt{Mineshige2007, Kawashima2012}), possibly with highly anisotropic radiation beamed toward us (e.g., \citealt{King2001}). 
Since the sources are extra Galactic, their mass donating companions have not been firmly identified except several cases \citep{Motch2014, Heida2015}, and hence little constraints have been obtained on the dynamical mass of their central objects.

In 2014, the $\sim1.4$~sec coherent hard X-ray pulsation was detected with \textit{NuSTAR} from a central region of the starburst galaxy M82 \citep{Bachetti2014}, and its source was identified by \textit{Chandra} as the ULX known as M82 X-2. 
After this epoch-making discovery, some authors \citep{Furst2016, Israel2017a, Israel2017b, Carpano2018} searched the archival X-ray data of ULXs for similar pulsations, and detected three more examples; NGC 7793 P-13 with a period of $\sim 0.4$~sec, NGC 5907 X-1 with $\sim 1.3$~sec, and NGC 300 ULX-1 with $\sim 32$ sec. 
These results have demonstrated that some ULXs do contain neutron stars (NSs) as their compact components, which are apparently radiating (if isotropic) at $\sim 100$ times the Eddington luminosity for a typical $1.4 \ M_{\rm{\odot}}$ NS. 
Although these ``NS ULXs'' impose a new challenge to our understanding of ULXs, they tend to exhibit spectra with higher cutoff energies and/or harder continua than those of the others \citep{Brightman2016a, Pintore2017}. 
Therefore, we still regard the remaining majority of ULXs as accreting BHs, and use the term ``ULX'' without including the NS ULXs.

Since observational information on ULXs is limited in other wavelengths, their X-ray spectra still provide major clues to their nature. As the luminosity changes, ULXs are considered to take three different spectral states with characteristic continuum shapes \citep{Gladstone2009, Miyawaki2009, Sutton2013}. 
One is so-called the Soft Power Law (SPL) state, wherein the spectrum exhibits a low energy hump at $\sim 1$~keV, on top of a soft Power Law (PL) component with a photon index of $\Gamma \sim 2.4$ and a high-energy roll over at $\sim 5$~keV. 
As the source brightens, it moves into a second state called the Hard PL (HPL) state, in which the PL continuum becomes literally harder ($\Gamma \sim 1.7$) and shows a slightly higher cutoff energy ($\sim 8$~keV) than in the SPL state. 
Finally, when the source brightens even further, the soft excess becomes less visible and the cutoff energy decreases, and the overall spectrum takes a convex shape; this is called the Multi-Color Disk like (MCD) state. 

These kinds of spectral transition phenomena are widely observed in other accreting compact objects, including black-hole binaries (BHBs; e.g., \citealt{Remillard2006}), low-mass X-ray binaries involving weakly magnetized NSs (e.g., \citealt{Sakurai2014}), and Seyfert galaxies as recently revealed by \citet{Noda2014}. 
In all these cases, the spectral states are approximately specified by the Eddington ratio $\eta \equiv L_{\rm{X}}/L_{\rm{edd}}$, where $L_{\rm{X}}$ is the X-ray luminosity. 
If this basic property of accreting systems also applies to ULXs, we are able to estimate how widely their masses are distributed, just by studying the scatter of state-transition threshold luminosities among ULXs.

While ULX spectra exhibit the various types of spectral continua as described above, they show little local features in their spectra such as emission/absorption lines or absorption edges. 
Especially, the features from Fe K-shell, which are ubiquitously seen in other accreting objects, are absent or very weak in their spectra. 
Although some authors succeeded to detect several resonance line features of Ne, O, and Fe from a few ULXs, these features are as weak as a few eV in equivalent width  \citep{Pinto2016, Walton2016}. 
Generally, these local features are considered to reflect the surrounding environment, including the geometry, density, velocity, and the ionization states of the accreting matters. 
Therefore, the absence or weakness of these features suggest some difference of ULXs from the other types of accreting objects in their accretion scheme. 

In the present paper, we analyze archival data of several representative ULX sources, and study the distribution of their state emerging luminosities by tracking their mid-term ($\sim10$~ks) to long-term ($>$days) variability. 
We also search their spectra for local features from Fe K-shell. 
Based on these observational results, a novel accretion scheme is proposed to explain the basic properties of ULXs. 

\section{Observations and Data Reductions}
We utilized archive data obtained with \textit{XMM-Newton}, \textit{Suzaku}, and \textit{NuSTAR}. 
To discriminate ULXs from less exotic accreting systems (including ordinary stellar-mass BHs), we limited our studies to those of which the observed maximum luminosity exceeds $L_{\rm{edd}}$ of the heaviest class of stellar-mass BHs in our galaxy ($\sim15 \ M_{\rm{\odot}}$), namely, $\sim2\times10^{39}$~erg~sec$^{-1}$. 
In addition, to acquire sufficient photon statistics, the targets were limited to those in nearby galaxies with a distance of $\le 5$~Mpc. Then, a typical ULX with $L_{\rm{X}}\sim2\times10^{39}$~erg~sec$^{-1}$ will yield a few thousand photons to these X-ray observatories in a typical exposure of several tens ksec. 
Finally, to study spectral variability, we further limited our sample objects to those which were observed more than twice with either of the three observatories. 
As a result, nine ULXs as listed in table \ref{tab:source_list} have been selected. Table \ref{tab:obs_list} summarizes basic information of the data sets we analyze in the present study.

\begin{table}
	\centering
	\caption{Summary of the ULXs studied in the present paper.}
	\label{tab:source_list}
	\begin{tabular}{cccc}
		\hline
		host galaxy   & $N^{\rm G}_{\rm H}$$^a$ & distance$^b$ & source name\\
				    & ($\times 10^{20}$ cm$^{-2}$)    & (Mpc)	     &		        	    \\
		\hline
		M33 		    &   4.4 		   	    & 0.88 	     & X-8		    \\
		Holmberg II  &   3.9 		   	    & 3.3	     & X-1		    \\
		IC 342	    &   34.5		   	    & 3.3	     & X-1		    \\
		Holmberg IX &   4.3		   	    & 3.4	     & X-1		    \\
		NGC 4190    &   2.1  		   	    & 3.5	     & X-1		    \\
		NGC 1313    &   3.6		   	    & 4.13	     & X-1		    \\
				    &			  	    & 		     & X-2		    \\
		M83		    & 	3.8	   	   	    & 4.61	     & ULX-1	    \\
				    &			 	    & 		     & ULX-2	    \\	
		\hline
	\end{tabular}
	\begin{flushleft}
	Notes:$^{a}$Galactic equivalent hydrogen column densities toward each source, taken from \citet{Dickey1990}. $^{b}$Distances toward the host galaxies taken from NASA/IPAC Extragalactic Database.
	\end{flushleft}
\end{table} 

\begin{table*}
	\caption{Observational log.}
	\label{tab:obs_list}
	\begin{tabular}{ccccccc}
	\hline\hline
	Host galaxy & ObsID$^{\rm a}$ & RA$^{\rm b}$ & Dec$^{\rm c}$ & Observation Start & Exposure$^{\rm d}$ & Observatory\\
	& & (deg) & (deg) &yyyy/mm/dd hh:mm:ss & (ks) &\\
	\hline\hline
	NGC 1313 & 0150280301 & 49.66304 & -66.59972 & 2003/12/21 01:54:45 & 7.4 & \textit{XMM-Newton}\\
			 & 0150280401 & 49.66196 & -66.60047 & 2003/12/23 04:50:43 & 10.6 & \textit{XMM-Newton}\\
			 & 0150280601 & 49.65758 & -66.60817 & 2004/01/08 03:30:34 & 6.0 & \textit{XMM-Newton}\\
	                  & 0150281101 & 49.65250 & -66.61319 & 2004/01/16 23:38:52 & 6.3 & \textit{XMM-Newton}\\
	                  & 0205230301 & 49.5170 & -66.60464 & 2004/06/05 06:08:51 & 8.5 & \textit{XMM-Newton}\\
	                  & 0205230401 & 49.55733 & -66.57292 & 2004/08/23 05:44:43 & 11.1 & \textit{XMM-Newton}\\
	                  & 0205230501 & 49.65583 & -66.58697 & 2004/11/23 06:59:33 & 12.3 & \textit{XMM-Newton}\\
	                  & 0205230601 & 49.63488 & -66.62122 & 2005/02/07 11:35:10 & 7.7 & \textit{XMM-Newton}\\
	                  & 100032010 & 49.5282 & -66.5443 & 2005/10/15 13:21:13 & 32.9 &\textit{Suzaku}\\
	                  & 0301860101 & 49.42263 & -66.57953 & 2006/03/06 16:43:12 & 16.7 & \textit{XMM-Newton}\\
	                  & 0405090101 & 49.59296 & -66.47497 & 2006/10/15 23:44:33 & 69.4 & \textit{XMM-Newton}\\
	                  & 703010010 & 49.6088 & -66.5434 & 2008/12/05 23:03:23 & 91.5 &\textit{Suzaku}\\
	                  & 30002035002 & 49.6268 & -66.5225 & 2012/12/16 13:56:07 & 100.8 &\textit{NuSTAR}\\
	                  & 0693850501$^{\dag}$ & 49.63108 & -66.49931 & 2012/12/16 16:00:25 & 92.1 & \textit{XMM-Newton}\\
	                  & 30002035004 & 49.6351 & -66.5268 & 2012/12/21 20:06:07 & 127.0 & \textit{NuSTAR}\\
	                  & 0693851201$^{\dag}$ & 49.63305 & -66.50272 & 2012/12/22 15:45:31 & 80.8 & \textit{XMM-Newton}\\
	                  & 0722650101 & 49.33767 & -66.55542 & 2013/06/08 05:21:44 & 11.3 & \textit{XMM-Newton}\\
	                  & 709023010 & 49.4840 & -66.5362 & 2014/05/27 05:41:09 & 107.3 &\textit{Suzaku}\\
	\hline
	Holmberg IX & 0111800101 & 148.89925 & 69.03797 & 2001/04/22 10:20:33 & 7.6 & \textit{XMM-Newton}\\
			    & 0112521001 & 149.46213 & 69.03264 & 2002/04/10 16:58:14 & 7.0 & \textit{XMM-Newton}\\
			    & 0112521101 & 149.46804 & 69.03222 & 2002/04/16 17:33:15 & 7.4 & \textit{XMM-Newton}\\
			    & 0200980101 & 149.50104 & 69.09072 & 2004/09/26 07:25:12 & 56.9 & \textit{XMM-Newton}\\
	                     & 0657801801 & 149.21079 & 69.09094 & 2011/09/26 04:17:58 & 11.9 & \textit{XMM-Newton}\\
	                     & 0693850801$^{\dag}$ & 149.47125 & 69.09203 & 2012/10/23 04:17:03 & 8.8 & \textit{XMM-Newton}\\
	                     & 0693851001$^{\dag}$ & 149.46817 & 69.09231 & 2012/10/27 04:04:42 & 3.3 & \textit{XMM-Newton}\\
	                     & 30002033006 & 151.1143 & 68.5532 & 2012/11/11 16:51:07 & 35.2 &\textit{NuSTAR}\\
	                     & 0693851701$^{\dag}$ & 149.44925 & 69.09156 & 2012/11/12 03:04:01 & 7.1 & \textit{XMM-Newton}\\
	                     & 30002033008 & 151.1297 & 68.5537 & 2012/11/14 01:31:07 & 14.5 & \textit{NuSTAR}\\
	                     & 0693851801$^{\dag}$ & 149.44767 & 69.09117 & 2012/11/14 02:55:07 & 6.3 & \textit{XMM-Newton}\\
	                     & 30002033010 & 151.1297 & 68.5537 & 2012/11/15 17:51:07 &  49.0 & \textit{NuSTAR}\\
	                     & 0693851101$^{\dag}$ & 149.44658 & 69.09083 & 2012/11/16 02:52:38 & 2.7 & \textit{XMM-Newton}\\
	\hline
	Holmberg II & 0112520601 & 124.8950 & 70.67828 & 2002/04/10 12:31:19 & 4.6 & \textit{XMM-Newton}\\
	                    & 0112520701 & 124.90204 & 70.67975 & 2002/04/16 12:06:26 & 3.9 & \textit{XMM-Newton}\\
	                    & 0112520901 & 124.8760 & 70.73356 & 2002/09/18 02:10:13 & 4.4 & \textit{XMM-Newton}\\
	                    & 0200470101 & 124.90079 & 70.67844 & 2004/04/15 20:08:43 & 33.8 & \textit{XMM-Newton}\\
	                    & 0561580401 & 124.87692 & 70.67714 & 2010/03/26 09:20:48 & 21.8 & \textit{XMM-Newton}\\
	                    & 30001031002 & 124.8517 & 70.684 & 2013/09/09 05:41:07 & 31.3 & \textit{NuSTAR}\\
	                    & 0724810101$^{\dag}$ & 124.88775 & 70.73364 & 2013/09/09 06:33:44 & 4.4 & \textit{XMM-Newton}\\
	                    & 30001031003 & 124.9776 & 70.693 & 2013/09/09 17:31:07 & 79.4 &\textit{NuSTAR}\\
	                    & 30001031005 & 124.9655 & 70.7048 & 2013/09/17 04:46:07& 111.1 &\textit{NuSTAR}\\
	                    & 0724810301$^{\dag}$ & 124.87921 & 70.73342 & 2013/09/17 06:03:28 & 4.9 & \textit{XMM-Newton}\\
	\hline
	IC 342 & 0206890101 & 56.57808 & 68.11403 & 2004/02/20 06:30:24 & 17.3 & \textit{XMM-Newton}\\
	            & 0206890201 & 56.47533 & 68.15733 & 2004/08/17 18:48:16 & 17.1 & \textit{XMM-Newton}\\
	            & 0206890401 & 56.57175 & 68.11256 & 2005/02/10 17:26:36 & 6.3 & \textit{XMM-Newton}\\
	            & 30002032002 & 58.4077 & 67.7677 & 2012/08/10 08:21:07 & 21.0 &\textit{NuSTAR}\\
	            & 0693850601$^{\dag}$ & 56.43046 & 68.10256 & 2012/08/11 20:06:44 & 35.5 & \textit{XMM-Newton}\\
	            & 30002032005 & 58.4635 & 67.844 & 2012/08/16 08:26:07 & 127.4 & \textit{NuSTAR}\\
	            & 0693851301$^{\dag}$ & 56.43146 & 68.10375 & 2012/08/17 19:48:41 & 32.4 & \textit{XMM-Newton}\\
	\hline
	M83 & 0723450101 & 204.2760 & -29.89686 & 2013/08/07 16:38:29 & 44.7 & \textit{XMM-Newton}\\
	        & 0723450201 & 204.26263 & -29.84075 & 2014/01/11 11:59:44 & 37.5 & \textit{XMM-Newton}\\
	        & 0729561201 & 204.28029 & -29.89628 & 2014/07/06 17:45:59 & 24.0 & \textit{XMM-Newton}\\
	        & 0729561001 & 204.26696 & -29.84067 & 2015/02/02 16:00:02 & 14.3 & \textit{XMM-Newton}\\
	\hline
	NGC 4190 & 0654650101 & 183.44208 & 36.60292 & 2010/06/06 12:08:27 & 1.9 & \textit{XMM-Newton}\\
	                  & 0654650201 & 183.44233 & 36.60336 & 2010/06/08 11:14:45 & 8.5 & \textit{XMM-Newton}\\
	                  & 0654650301 & 183.43221 & 36.65992 & 2010/11/25 01:24:51 & 11.2 & \textit{XMM-Newton}\\
	\hline
	\end{tabular}
	\begin{flushleft}
	{\footnotesize
		Notes: $^{\dag}$Observations that are combined with \textit{NuSTAR}. $^{\rm a}$Observational ID numbers of the data sets. $^{\rm b}$Right-ascensions of the aim points. $^{\rm c}$Declination of the telescope aim points. $^{\rm d}$Exposure time after the reduction of the SAA-passage/proton-flare intervals.\\
	}
	\end{flushleft}
\end{table*}

\begin{table*}
	\contcaption{}
	\label{tab:obs_list2}
	\begin{tabular}{ccccccc}
	\hline\hline
	Host galaxy & ObsID$^{\rm{a}}$ & RA$^{\rm{b}}$ & Dec$^{\rm{c}}$ & Observation Start & Exposure$^{\rm{d}}$ & Observatory\\
	& & (deg) & (deg) &yyyy/mm/dd hh:mm:ss & (ks) &\\
	\hline\hline

	M33 & 0102640101 & 23.43317 & 30.67694 & 2000/08/04 05:17:12 & 5.3 & \textit{XMM-Newton}\\
	        & 0102640301 & 23.35442 & 30.88608 & 2000/08/07 01:03:49 & 3.5 & \textit{XMM-Newton}\\
	        & 0141980801 & 23.49225 & 30.64731 & 2003/02/12 15:17:46 & 6.8 & \textit{XMM-Newton}\\
	        &704016010 & 23.4896 & 30.5679 & 2010/01/11 01:47:07 & 82.5 & \textit{Suzaku}\\
	\hline
	\hline
	\end{tabular}
	\begin{flushleft}
	{\footnotesize
		Notes: $^{\dag}$Observations that are combined with \textit{NuSTAR}. $^{\rm{a}}$Observational ID numbers of the data sets. $^{\rm{b}}$Right-ascensions of the aim points. $^{\rm{c}}$Declination of the telescope aim points. $^{\rm d}$Exposure time after the reduction of the SAA-passage/proton-flare intervals.\\
	}
	\end{flushleft}
\end{table*}

\subsection{\textit{Suzaku} data}
The \textit{Suzaku} satellite \citep{Mitsuda2007} is capable of broad band spectroscopy, by combining the X-ray Imaging Spectrometer (XIS; \citealt{Koyama2007}) operating in the $0.3\--12$~keV band, and the Hard X-ray Detector (HXD; \citealt{Takahashi2007}) in $15\--600$~keV. 
However, we utilize only the XIS data, since those of the HXD, being a non-imaging instrument, are often contaminated by emission from other X-ray sources.
Because one of the four XIS cameras lost its function since 2006, we utilized the remaining two front Illuminated (FI) type CCD cameras (XIS0, XIS3) and one back illuminated (BI) type one (XIS1). 
In spectral analysis, we excluded the $1.5\--2.0$~keV band to avoid calibration uncertainties therein, and co-added data from the two FI-XIS cameras because they have nearly identical responses. 

To minimize the background, we discarded the data taken during and within 436~sec after exiting the South Atlantic Anomaly, and when the spacecraft was in the region where the cut off rigidity is less than 6~GV. 
Furthermore, to avoid strong solar X-ray albedo from the Earth's atmosphere, we set the satellite elevation to be higher than $20^{\circ}$ from the sunlit Earth. 
Incorporating these data screenings, the data reduction was carried out using the latest calibration database files and the $\tt{aepipeline}$ software, which is included in High Energy Astronomy SOFT (HEASOFT).

Each ULX spectrum was accumulated over a circular region with a radius of $2'.5$, which corresponds to $\sim90\%$ of the encircled energy function of the X-ray Telescope.
Background spectra were extracted from circular or annular regions wherein no apparent X-ray source is present. In order to maximize the signal-to-noise ratio, the background accumulation region was set to be relatively large, with a radius up to $2'$.

\subsection{\textit{XMM-Newton} data}
The \textit{XMM-Newton} observatory \citep{Barre1999} allows high-sensitivity observations in $0.3\--10$~keV with the European Photon Imaging Cameras (EPIC), which consists of two Metal-Oxide Semiconductor type CCD detectors (MOS; \citealt{Turner2001}), and a pn junction type one (PN; \citealt{Struder2001}).

The gain calibrations and removal of bad quality events for MOS and PN were carried out with  {\tt emchain} and {\tt epchain}, respectively, both included in the Science Analysis System package. 
In addition, solar flare proton events were also removed by referring to light curves above $10$~keV, where the X-ray mirror has nearly zero reflectivity. We discarded events taken in the time intervals where the count rate is twice as high as the average.We created the MOS-1 and MOS-2 spectra separately, and fitted them simultaneously. 

Considering the angular resolution of \textit{XMM-Newton} ($\sim 15''$ in half power diameter), we extracted spectra of the target ULXs from circular regions with a radius of $30''$, which encircles typically $83\--88\%$ of the source photons. The background spectra were taken from the same observation as the on-source spectra, but from a region wherein no apparent X-ray sources are detected. To avoid known positional dependence, we further limited the background region to be within the CCD chip in which the target source is present. As a result, the background region generally became a circle with a radius of $30''\--45''$ and $30''\--60''$ for PN and MOS, respectively.

\subsection{\textit{NuSTAR} data}
The \textit{NuSTAR} satellite \citep{Harrison2013} is capable of imaging spectroscopy in the $3\--80$~keV band, with two CdZnTe detector modules called Focal Plane Modules A and B, to be abbreviated as FPM-A and FPM-B, respectively \citep{Kitaguchi2014}. 
In the present paper, we utilize \textit{NuSTAR} data for spectroscopic studies of ULXs in a broader energy band than those of the other satellites.

For the same reason as in \textit{Suzaku}, we discarded the data taken during the South Atlantic Anomaly, or when the Earth was in the field of view, or when the elevation angle of the telescope above the Earth's limb was less than $3^{\circ}$. 
We also excluded the data taken when the source made an excursion by $> 6'$ from the nominal position due to wobbles of the optical bench. 
All these data reductions and other calibrations were performed with the software tools, {\tt nupipeline} and {\tt nuproducts}, included in HEASOFT.

The spectra of ULXs were extracted from circular regions of $30''$ radius, while those of background from regions with a radius of $2'$. Since a single FPM module consists of four CdZnTe detector chips, we defined the background region in the same chip as the target source, for the same reason as in \textit{XMM-Newton}.
Just like the EPIC data, we created the FPMA and FPMB spectra separately, and fitted them simultaneously.

\section{Data Analysis and Results}
\subsection{Light Curves}
\begin{figure*}
 \includegraphics[width=\textwidth]{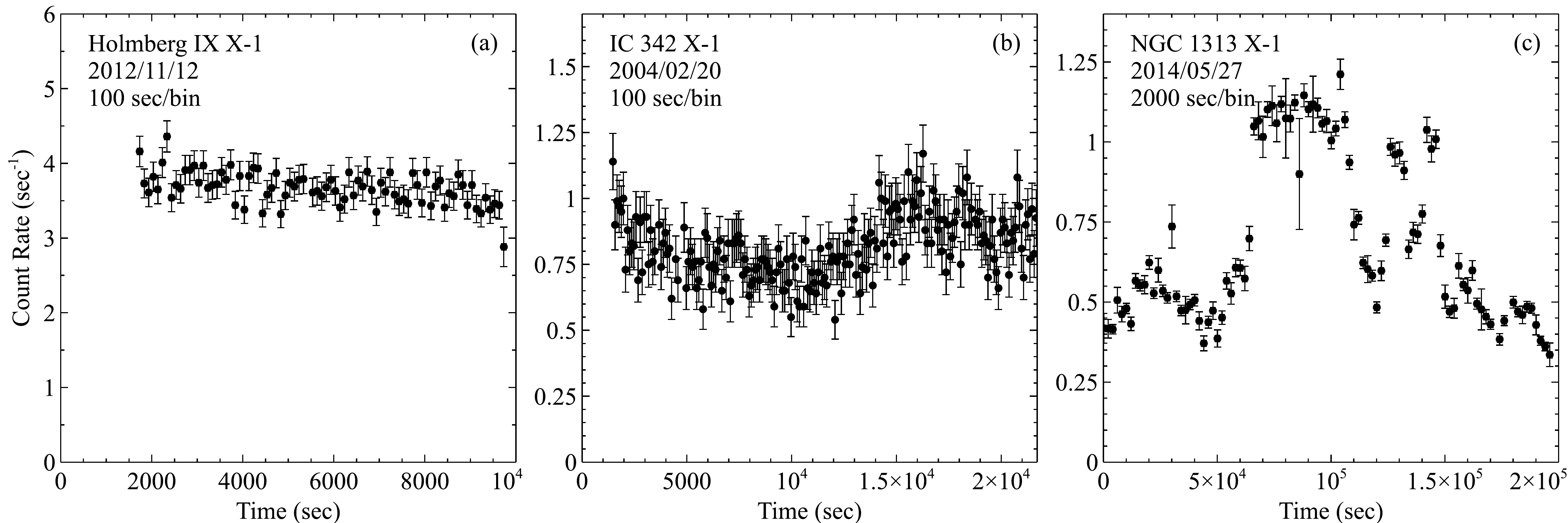}
 \caption{Examples of $0.3\--10$~keV background-subtracted light curves of the objects in the present sample. The source name and the observing epoch are indicated at the top of each panel. The data in (a) and (b) were taken with EPIC-PN, while that of (c) with XIS0+XIS3. 
 They are uncorrected for the vignetting effect. 
 The zero second of each light curve is set to the start point of its observation. 
 All events taken during high background rates (e.g., soft proton flares) were excluded.}
 \label{fig:lightcurve}
 \end{figure*}
In figure \ref{fig:lightcurve}, we present background subtracted $0.3\--10$~keV light curves obtained from three representative observations in our sample; that of Holmberg IX X-1, IC342 X-1, and NGC 1313 X-1. 
As seen in the first two sources, typical X-ray intensity variations of the sample ULXs were no more than $30 \%$ within each observation. 
Hence, we basically derived a single averaged spectrum from each data set. However, as in figure \ref{fig:lightcurve} c, NGC 1313 X-1 varied, on 2014 May 27, by a factor of $\sim 2$ (in $0.5\--10$~keV) on a time scale of several ks. We hence divided this particular data set into three intensity intervals, low ($< 0.65$ counts sec$^{-1}$), middle ($0.65\--0.85$ counts sec$^{-1}$), and high ($> 0.85$ counts sec$^{-1}$), and made a spectrum from each of them. 

\subsection{Continuum Spectroscopy}
\subsubsection{The model fitting}
The obtained spectra are presented in figures \ref{fig:1313_res1} and \ref{fig:fit_res1}, after subtracting the background and using the $\nu F\nu$ form where the instrumental responses are approximately removed. 
Thus, most of the sources exhibited significant changes in their spectral shapes over the observational history spanning $\sim10$ years. 
In many of them, the variation is more pronounced in $> 1$~keV than in $< 1$~keV. 
Some sources, including NGC 1313 X-1, Holmberg IX X-1, IC 342 X-1, and M83 ULXs, showed series of convex spectra which are typical of the MCD state. 
As they became dimmer, the spectrum started showing the hard PL continua and soft excess components at $\sim 1$~keV, which are typical characteristics of the HPL state. 
Furthermore, in NGC 1313 X-1 and Holmberg II X-1, the PL slope softened to $\Gamma >2.0$ and the soft excess became more prominent as the luminosity decreased; the spectra finally reached the SPL state. This behavior is consistent with that found, e.g., by \citet{Gladstone2009}, \citet{Sutton2013}.

To quantify the shapes of the sample spectra, we fitted them with a model commonly used in studies of accreting BH systems; it consists of a multi-color disk model and its thermal Comptonization (MCD+THC modeling; \citealt{Makishima2008}). 
In the modeling, we assume that a standard accretion disk \citep{Shakura1973} is partially covered by a hot electron cloud, which Compton up-scatters some fraction of the photons from the disk \citep{Gladstone2009}. 
The remaining disk photons are assumed to reach the observer as the direct MCD component.
According to \citet{Gladstone2009}, this model can describe the three spectral states in a unified way, and provide reasonable physical interpretations. We retain the two components even if either of them is not strongly required by the data, because its upper limit is still meaningful.

Although the MCD+THC modeling is widely used to fit spectra in the HPL and SPL states, those in the MCD state, with convex shapes and high luminosity, are often explained with an alternative accretion disk model called ``slim disk''. 
This is an accretion-flow solution expected to emerge near or above the Eddington rate \citep{Abramowicz1988, Watarai2000}, and characterized by a flatter radial temperature gradient than the standard disk. 
A slim-disk spectrum is often approximated analytically with a ``free-p disk'' model, which is a modified MCD model with the temperature gradient index $p$ set free, instead of being fixed at $p=0.75$ as in a standard disk.
In fact, the slim disk model reproduces the same MCD state spectra as successfully as the MCD+THC modeling (e.g., \citealt{Miyawaki2009, Middleton2011, Bachetti2013}), usually without significant statistical differences. 
However, \citet{Miyawaki2009} have showed that the slim-disk interpretation of the spectrum of M82 X-1 in the MCD state leads to some physical inconsistency, including too high a disk temperature. 
Furthermore, the model also fails to give a consistent explanation to the spectral variability of several ULXs (e.g., \citealt{Mizuno2007}). 
For these reasons, we employ the MCD+THC modeling throughout the present work.

Using the above model, we analyze background-subtracted spectra of the sample ULXs with the HEASOFT XSPEC analysis software, employing the chi-squared statistics.
Prior to the fitting, the background-subtracted spectra were re-binned, so that each energy bin has $\gtrsim 20$ counts.
Although the \textit{NuSTAR} FPM is capable of detecting X-rays up to $\sim 80$ keV, the curving spectral shapes and rebinning criteria (typically $\delta E/E\sim0.1$ and $\gtrsim 5$ in signal-to-noise ratio) allowed us to detect the sources only up to $\sim 15$~keV.
The {\tt diskbb} model \citep{Mitsuda1984, Makishima1986} and the {\tt nthcomp} model \citep{Zdziarski1996, Zycki1999} were employed to express the direct MCD and the THC components, respectively. 
The former has the innermost disk radius $R_{\rm{raw}}$ and the temperature $T_{\rm{in}}$ as free parameters. 
The latter is specified by the photon index $\Gamma$ describing the continuum slope, and the electron temperature $T_{\rm{e}}$ characterizing the high-energy bend. 
The seed photon temperature of {\tt nthcomp} is set to be identical to $T_{\rm{in}}$. 
Assuming that the photon number is conserved in Compton scattering, we can convert the {\tt nthcomp} flux to the linear size of the disk, $R_{\rm{thc}}$, that is covered with the corona. The conversion is made by calculating a {\tt diskbb} model that gives the same temperature and photon flux as the {\tt nthcomp} component.
Accordingly, the innermost radius of the whole accretion disk $R_{\rm{tot}}$ can be calculated as
\begin{equation}
R_{\rm{tot}}^{2} = R_{\rm{raw}}^{2} + R_{\rm{thc}}^{2}
\label{eq:rtot}
\end{equation}
(e.g., \citealt{Kubota2004}, \citealt{Makishima2008}).

The photoelectric absorption was taken into account by applying two {\tt tbabs} \citep{Wilms2000} factors to the constructed model, one representing the Galactic line-of sight absorption and the other that from the host galaxy (including circum-source contributions).
The equivalent hydrogen column density $N^{G}_{\rm H}$ of the former was fixed at the values by \citet{Dickey1990}, whereas that of the latter, to be denoted as $N_{\rm H}$, was left free.
An exception is NGC 1313 X-1, where we replaced the second {\tt tbabs} to the {\tt vphabs} model, which allows us to freely change the abundances of individual elements. This is because the source is possibly residing in an oxygen-poor environment \citep{Mizuno2007}; following \citet{Mizuno2007}, we fixed the abundance of oxygen to 0.68 times the solar value, while the others at the solar abundance.

Finally, we multiplied an additional constant factor to the model, to deal with systematic relative uncertainties in the absolute photoelectric sensitivity among different types of detectors. This parameter was allowed to vary freely when the spectral analysis involve different instruments; otherwise, it was frozen to unity.

Table \ref{tab:fitparam} summarizes the best-fit parameters and the fit goodness, obtained through the spectral fitting analysis described above.  
Here and hereafter, all the errors refer to $90\%$ confidence level unless stated otherwise. 
Although most of the sources showed the strong variability in their spectral shapes, the MCD+THC model has successfully reproduced most of the spectra, as shown by solid lines in figures \ref{fig:1313_res1} and \ref{fig:fit_res1}.
In agreement with the low variability below 1~keV, $N_{\rm{H}}$ was relatively stable, and rather low as $N_{\rm{H}}\sim 10^{21}$~cm$^{-2}$ or even less in some sources (e.g., M83 ULX-1 and ULX-2). 
For example, the X-ray luminosity of NGC 1313 X-1 changed nearly by an order of magnitude, but the error-weighted mean and the standard deviation of the column density was $N_{\rm{H}} = (2.1\pm0.9)\times10^{21}$~cm$^{-2}$.

Due to the characteristic spectral bend at $5\--7$~keV, all sources required rather low electron temperatures as $T_{\rm{e}} = 0.9\--2.0$~keV, except data with insufficient statistics in the $7\-- 10$~keV band. 
The derived photon index is distributed over a rather wide range of $\Gamma = 1.6\-- 2.7$, reflecting the large spectral shape changes above 2~keV. Since $\Gamma$ represents the Compton y-parameter as $y=4/[(0.5+\Gamma)^2-2.5]$, we can obtain the optical depth of the corona, $\tau$, by combing $\Gamma$ and $T_{\rm{e}}$ as
\begin{equation}
\tau = \left[2.25 + \frac{3}{(T_{\rm{e}}/511 {\rm{\ keV}})[(\Gamma + 0.5)^{2} - 2.25]}\right]^{0.5} - \frac{3}{2}
\label{eq:tau}
\end{equation}
\citep{Sunyaev1980}. 
This equation generally yields $\tau \sim 5\--19$ for the present sample (table \ref{tab:fitparam}). 
Since $\tau$ depends on the two parameters, $T_{\rm e}$ and $\Gamma$, which are correlated with each other, the statistical uncertainty of $\tau$ was calculated by taking the covariance between them.

Although the MCD+THC model has been generally successful, a few spectra of NGC 1313 X-1 gave slightly worse fits as $\chi^2/\nu=1.25\--1.35$. 
These are all caused by two types of local features seen in residuals below 1~keV. 
One is the residuals at 0.56~keV (e.g., figure \ref{fig:1313_res1} panels A and B) which is likely to be neutral Oxygen K-edge absorption feature caused by inaccurate estimates of the Oxygen abundance of local absorbers around NGC 1313 X-1. 
In fact, the residual disappears if we let the Oxygen abundance in {\tt vphabs} increase to 0.71 times solar, from the tentatively assumed value of 0.68.
The other is an unaccounted positive structure at $\sim 1$~keV (e.g., figure \ref{fig:1313_res1} panel C). 
Similar local residuals are also reported in other ULXs (e.g., NGC 5408 X-1; \citealt{Middleton2014}), and their origin is still under a debate. 
Further examination of the $\sim1$~keV residuals is beyond the scope of the present paper, because our major objective is to quantify the continuum.

\subsubsection{Spectral state characterization}
Now that the MCD+THC modeling has thus been successful, next we try to characterize the individual spectral states in terms of the obtained model parameters. 
For example, as shown by figure \ref{fig:fit_res1} (A) and (B), Holmberg IX X-1 exhibited two distinct types of spectral states, to be identified with the HPL state and the MCD state. 
As the source brightened, the spectrum changed gradually in shape from two-humped ones (figure \ref{fig:fit_res1} panel B) typical of the former state, to more convex ones (figure \ref{fig:fit_res1} panel A) often seen in the latter state. 
In terms of the spectral fit parameters, this spectral change, or transition, from the HPL to MCD states as the luminosity increases, can be characterized by two effects. 
One is mutual approach between $T_{\rm{in}}$ and $T_{\rm{e}}$, because $T_{\rm{in}}$ increased from $\sim0.3$~keV to $0.6\--0.9$~keV, whereas $T_{\rm{e}}$ decreased slightly from $2.6$~keV to $1.6\--2.0$~keV.
The other is a marked decrease of $R_{\rm{raw}}$, or the directly visible disk fraction, because the soft excess becomes no longer visible in the convex-shaped MCD state.

\begin{figure*}
	\includegraphics[width=\textwidth]{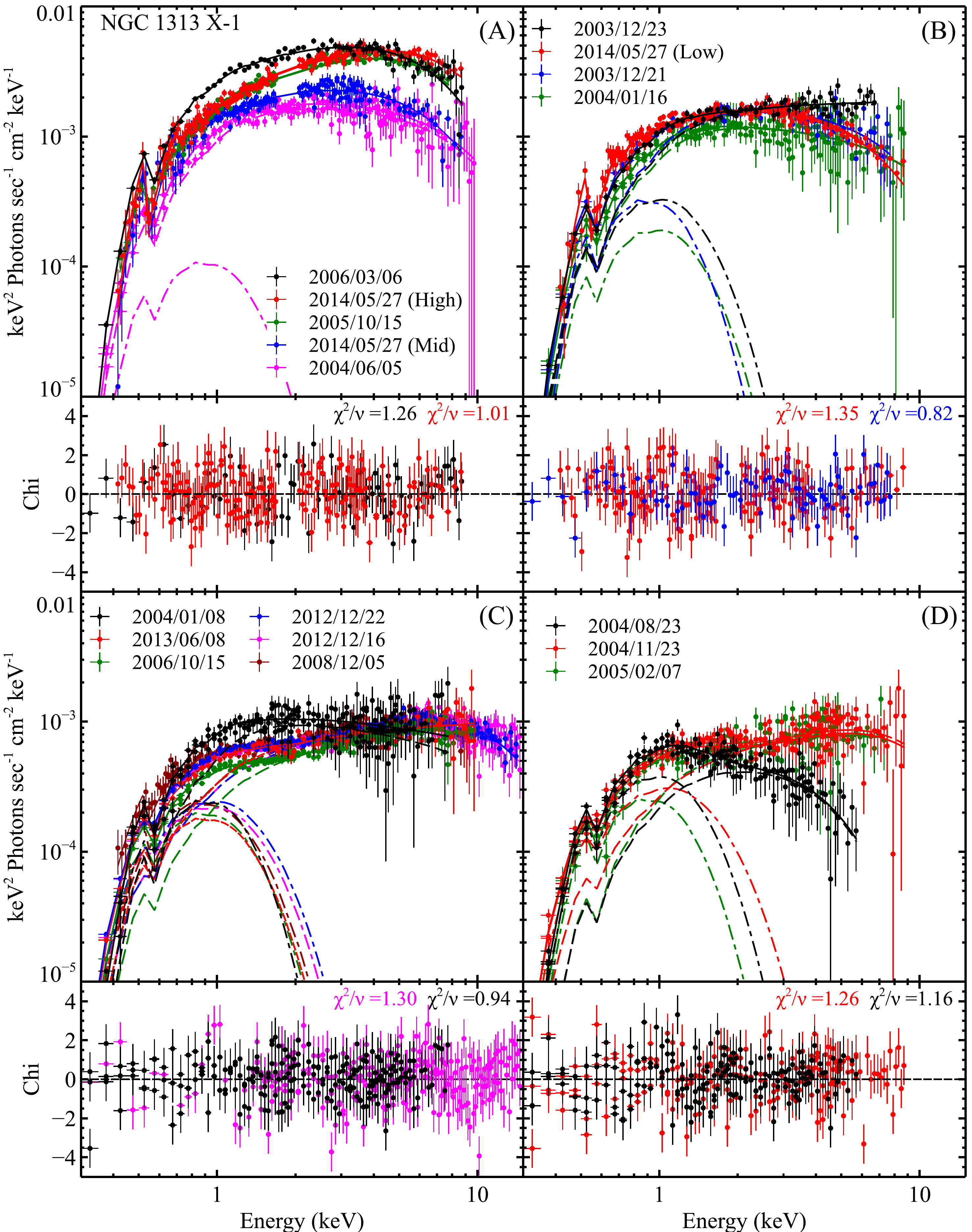}
	\caption{Spectra of NGC 1313 X-1 unfolded with the best fit MCD+THC models (solid lines). The dot-dashed lines represent the MCD component, while the dashed ones are the THC component. They are grouped into four in terms of their $0.3\--10$~keV luminosity. (A): $L_{\rm{X}} > 1.3\times10^{40}$~erg~sec$^{-1}$. (B): $8\times10^{39}$~erg~sec$^{-1}$ $<\ L_{\rm{X}}\ < 1.3\times10^{40}$~erg~sec$^{-1}$. (C): $5\times10^{39}$~erg~sec$^{-1}$ $<\ L_{\rm{X}}\ < 8\times 10^{39}$~erg~sec$^{-1}$. (D): $L_{\rm{X}} < 6.4\times 10^{39}$~erg~sec$^{-1}$. Each elongated panel presented at the bottom shows the residuals and reduced chi-squared values obtained from the best/worst fit results in the respective spectral group.}
	\label{fig:1313_res1}
\end{figure*}

Spectral transitions of the same type as that of Holmberg IX X-1 were also seen in other five sources, NGC 1313 X-1, IC 342 X-1, M83 ULX-1, M83 ULX-2, and NGC 4190 X-1 (see figures \ref{fig:1313_res1} and \ref{fig:fit_res1}). 
Most of them also showed increasing $T_{\rm{in}}$, decreasing $R_{\rm{raw}}$, and decreasing or comparable $T_{\rm{e}}$ through their transitions from the HPL states to the MCD states.
As an exception, several data sets of NGC 1313 X-1 and IC 342 X-1 in the MCD state showed rather lower temperatures as $T_{\rm{in}} = 0.13\--0.2$~keV and extremely large radii as $R_{\rm{raw}}=15000\--17000$~km, despite their convex shaped spectra which are similar to those of the others. 
This is rather unphysical, because the results would imply that extremely large disks are present in the MCD states, and most of their emission is hidden below the detectable energy band. 
Therefore, we applied a slight modification to our modeling, as detailed in Appendix, assuming that the corona covers only a limited inner region of the disk.
As shown in table \ref{tab:fitparam}, the alternative model gave similarly good fit as the original model, or even slightly better. 
The innermost temperature of the corona-covered disk region, $T_{\rm{in2}} = 0.6\--0.9$~keV, is consistent with the results obtained in the previous study \citep{Kobayashi2016}, and is higher than that in the HPL state ($\sim0.2$ keV). 
Thus, in these two sources, the transitions from the HPL state to the MCD state are also successfully explained by a decrease in the $T_{\rm e}/T_{\rm in}$ ratio, and an increase of the covering fraction of the corona above the accretion disk.

While most of the sources resided either in the HPL or the MCD states, Holmberg II X-1 and NGC 1313 X-1 occasionally exhibited, in addition to the HPL spectra, those with even softer ($\Gamma > 2.0$) continuum, which is typical of the SPL state.
This state assignment agrees with previous observations (e.g., \citealt{Feng2006, Kajava2009, Kajava2012}). 
The spectrum of Holmberg II X-1 obtained on 2002 September 18 (shown in green in figure \ref{fig:fit_res1} panel D) exhibits a strong hump with $T_{\rm{in}}=0.16\pm0.06$~keV and a soft $\Gamma > 2.6$ PL continuum with the lowest coronal temperature of $T_{\rm{e}} = 0.8^{+1.0}_{-0.3}$~keV. 
As the source became more luminous, the spectrum made a transition toward the HPL state; the PL continuum hardened to $\Gamma = 1.8$ and $T_{\rm{e}}$ increased to $\sim2.4$~keV, while $T_{\rm{in}}$ remained nearly unchanged at $\sim 0.2$~keV. 
Thus, the evolution from the SPL to the HPL state can be characterized not only by a hardening of the continuum, but also by an increase of the difference between $T_{\rm{in}}$ and $T_{\rm{e}}$. 

The only SPL-state data set from NGC 1313 X-1 (2004 August 23; figure \ref{fig:1313_res1} panel D) also showed the lowest electron temperature of $T_{\rm{e}} = 0.9^{+0.6}_{-0.2}$, and a strong soft excess which can be reproduced by a disk with $T_{\rm{in}}=0.22\pm0.05$~keV.  
Just like the case of Holmberg II X-1, $T_{\rm{e}}$ significantly increased up to $\sim 2.6$~keV as the source brightened and made a transition to the HPL states, while $T_{\rm{in}}$ were nearly constant at $0.2$~keV. 
Thus, both NGC 1313 X-1 and Holmberg II X-1 showed an increase in the $T_{\rm e}/T_{\rm in}$ ratio through their transitions from the SPL state to the HPL state, while $T_{\rm{in}}$ remained nearly unchanged.

In contrast to the above seven sources, the remaining two, M33 X-8 and NGC 1313 X-2, exhibited the convex MCD state spectra throughout all epochs (figure \ref{fig:fit_res1} panels F and H). Both sources yielded rather high $T_{\rm in}$ as $0.21\--0.53$~keV and low $T_{\rm e}$ as $1.1\--2.2$~keV, compared to those seen in the HPL states of the other sources. 
In addition, the directly visible disk component was hardly required because these convex spectra were individually reproduced almost solely by an {\tt nthcomp} model with a high $T_{\rm{in}}$ and low $T_{\rm{e}}$. 
These characteristics in the derived parameters are consistent with those in the MCD states of the other sources described so far. 

\begin{figure*}
	\includegraphics[width=\textwidth]{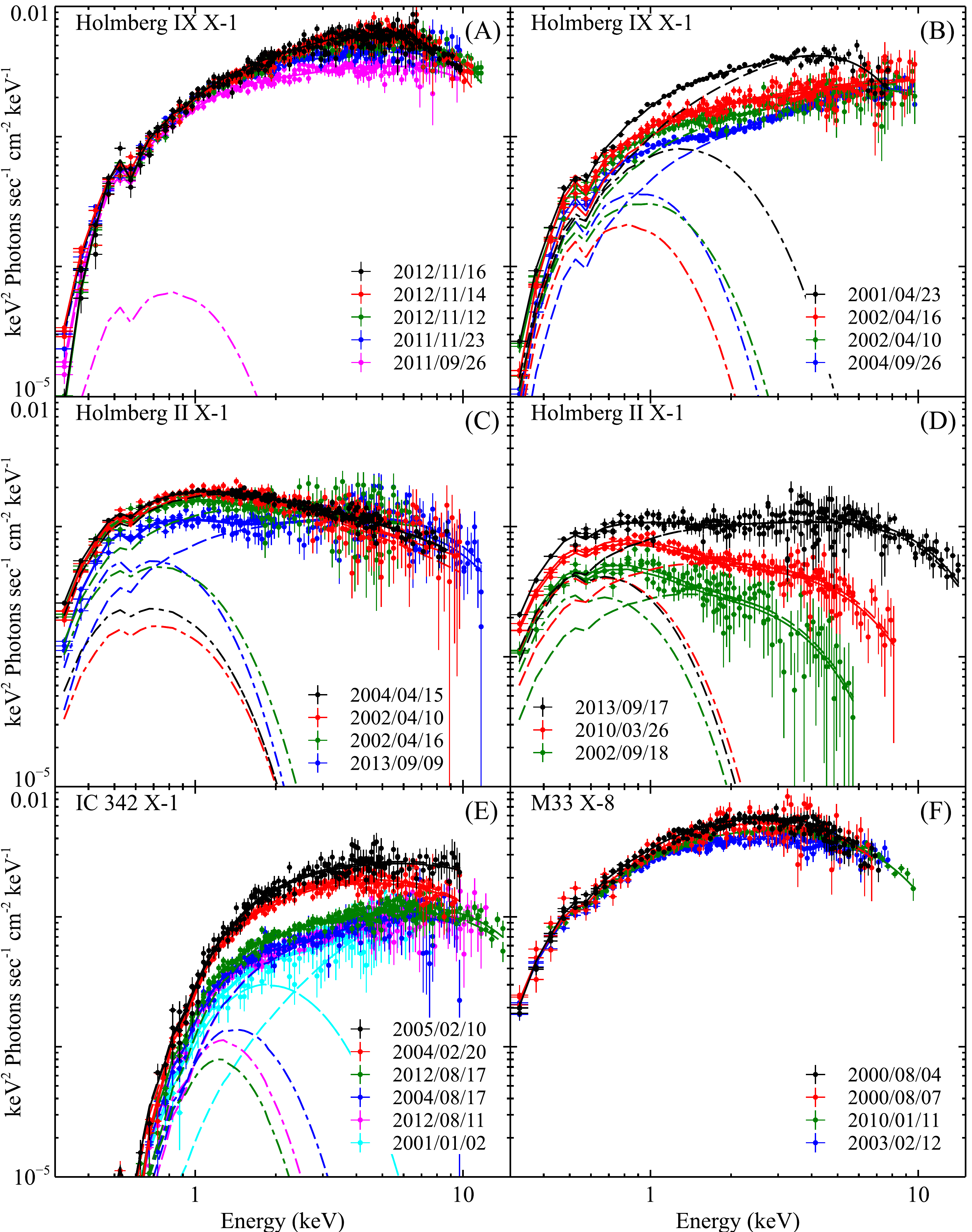}
	\caption{Spectra obtained from other four ULXs, presented in the same way as those of NGC 1313 X-1. For clarity, the spectra of Holmberg IX X-1 and Holmberg II X-1 are divided at the luminosity of $1.5\times10^{40}$ erg sec$^{-1}$ and $5\times 10^{39}$ erg sec$^{-1}$, respectively.}
	\label{fig:fit_res1}
\end{figure*}

\begin{figure*}
	\includegraphics[width=\textwidth]{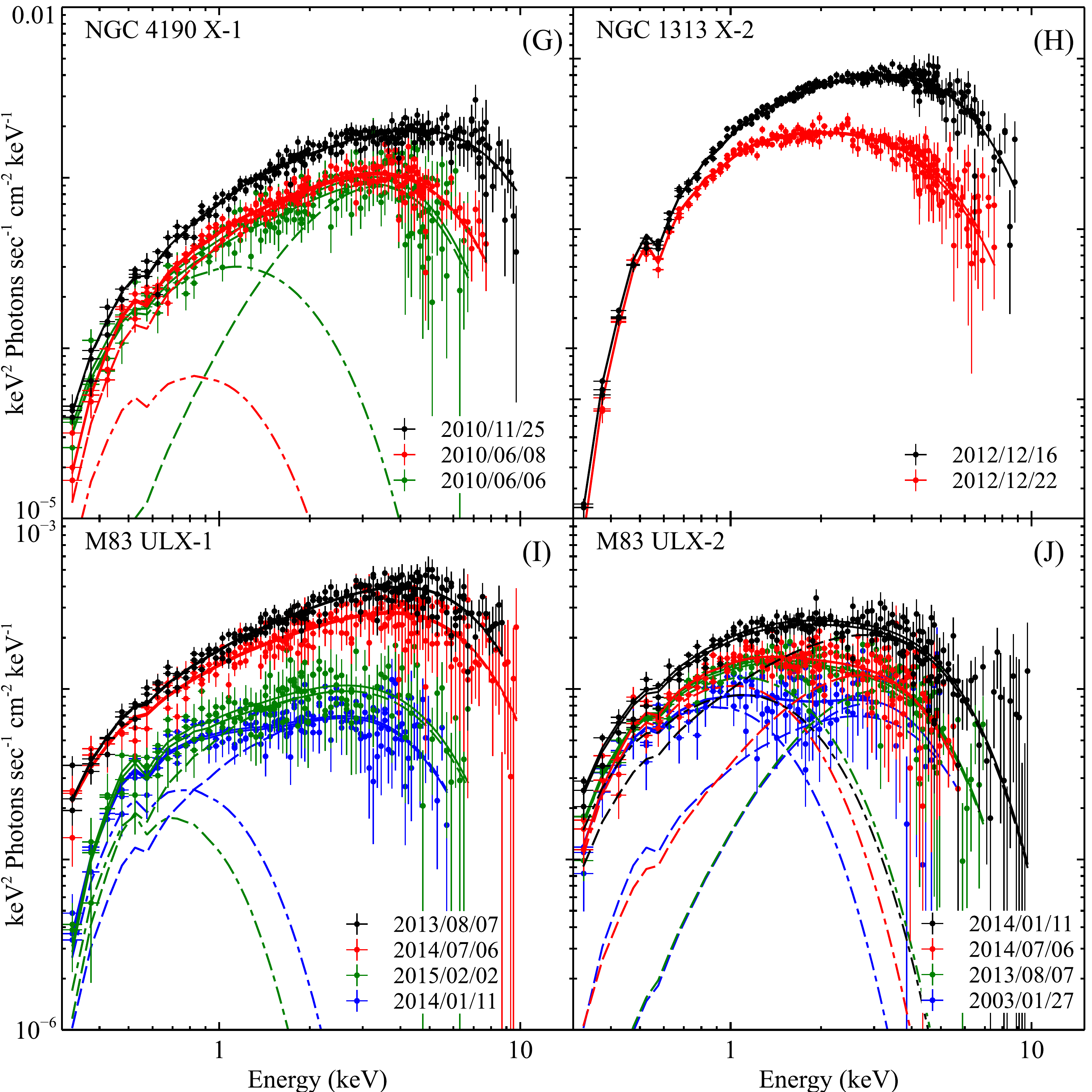}
	\contcaption{Spectra obtained from the remaining four sources.}
\end{figure*}

		\begin{table*}
		\caption{Parameters obtained from the spectral fitting.}
		\label{tab:fitparam}
		\begin{tabular}{cccccccccccc}
		\hline
		\hline
		Date$^{\rm{a}}$ & $N_{\rm{H}}$$^{\rm{b}}$ & $T_{\rm{e}}$ & $\Gamma$ & $\tau$$^{\rm{c}}$ & $T_{\rm{in}}$ or $T_{\rm{in1}}$$^{\rm{d}}$ & $R_{\rm{raw}}$$^{\rm{e}}$ or & $T_{\rm{in2}}$$^{\rm{j}}$ & $R_{\rm{tot}}$$^{\rm{f}}$ & $L_{\rm{disk}}$$^{\rm{g}}$ & $L_{\rm{X}}$$^{\rm{h}}$ & $\chi^{2}/\nu$ ($\nu$)\\
		YY/MM/DD& & (keV) & & & (keV) & $R_{\rm{raw1}}$$^{\rm{i}}$ & (keV) & & & & \\
		\hline\hline
		\multicolumn{12}{c}{NGC 1313 X-1}\\
		 03/12/21& $3.2^{+1.3}_{-0.9}$ & $1.8^{+1.6}_{-0.4}$ & $2.1\pm0.2$ & 12 & $0.18^{+0.08}_{-0.04}$ & $7.4^{+18.0}_{-5.2}$ & -- & $9.4^{+14.4}_{-4.1}$ & 12.0 & 10.8 & 0.82 (72)\\
		 03/12/23& $2.7^{+0.7}_{-0.6}$ & $> 1.9$ & $1.96^{+0.08}_{-0.06}$ &  $> 11$ & $0.22^{+0.07}_{-0.05}$ & $3.3^{+4.3}_{-1.8}$ & -- & $4.6^{+3.1}_{-1.3}$ & 6.4 & 13.8 & 0.93 (67)\\
		 04/01/08& $3\pm1$ & $> 2.6$  & $2.31^{+0.09}_{-0.34}$ & $> 8$ & $0.18^{+0.08}_{-0.03}$ & $7.0^{+11}_{-5}$ & -- &  $8.7^{+8.8}_{-4.0}$ & 10.2 & 7.5 & 0.94 (207)\\
		 04/01/16& $2.8^{+1.0}_{-0.7}$ & $2.6^{+2.6}_{-1.1}$ & $2.2^{+0.2}_{-0.3}$ & 9.3 & $0.21^{+0.12}_{-0.06}$ & $3.1^{+9.0}_{-1.9}$ & -- & $4.7^{+5.9}_{-1.2}$ & 5.5 & 8.4 & 1.12 (212)\\
		 04/06/05& $2.6^{+0.8}_{-0.6}$ & $2.0^{+0.7}_{-0.4}$ & $2.1\pm0.1$ & 11 & $0.19^{+0.12}_{-0.04}$ & $< 8.6$ & -- & $5.6^{+4.4}_{-1.5}$ & 5.3 & 11.4 & 1.05 (227)\\
		 04/08/23& $2.8^{+0.6}_{-0.5}$ & $0.9^{+0.6}_{-0.2}$ & $2.0^{+0.5}_{-0.8}$ & 19 & $0.22^{+0.05}_{-0.04}$ & $4.0^{+3.2}_{-1.5}$ & -- & $4.4^{+3.0}_{-1.4}$ & 5.9 & 3.3 & 1.16 (167)\\
		 04/11/23& $1.7^{+0.5}_{-0.4}$ & $2.0^{+8.2}_{-0.6}$ & $1.7^{+0.2}_{-0.3}$ & 16 & $0.29^{+0.08}_{-0.06}$ & $1.4^{+0.8}_{-0.5}$ & -- & $1.8^{+0.6}_{-0.4}$ & 3.0 & 6.4 & 1.26 (212)\\
		 05/02/07& $3.0^{+1.0}_{-0.8}$ & $3^{+3}_{-1}$ & $1.9^{+0.1}_{-0.2}$ & 11 & $0.185^{+0.05}_{-0.03}$ &  $6^{+7}_{-3}$ & -- & $6.3^{+6.3}_{-2.7}$ & 5 & 5.8 & 1.18 (195)\\
		 05/10/15& $2.9^{+0.8}_{-0.9}$ & $1.6\pm0.1$ & $1.75\pm0.05$ & 17 & $0.17^{+0.05}_{-0.02}$ & $7.2^{+8.4}_{-5.4}$ & -- & $10^{13}_{-9}$ & 10.6 & 24.7 & 1.35 (127)\\
		 		& $2.6\pm0.5$ & $> 2.7$ & $2.7^{+0.2}_{-0.6}$ & $> 7$ & $0.22\pm0.02$ & $4.5^{+5.4}_{-2.5}$ & $0.9^{+0.1}_{-0.2}$ & $0.30\pm0.01$ & 8 & 24.7 & 1.23 (135)\\		 
		 06/03/06& $3.2^{+0.8}_{-0.7}$ & $1.5^{+0.2}_{-0.1}$ & $1.92\pm0.09$ & 15 & $0.17^{+0.04}_{-0.03}$ & $12.0^{+14.0}_{-7.0}$ & -- & $15^{+6}_{-3}$ & 24.3 & 31.8 & 1.41 (77)\\
		 		& $2.7^{+0.4}_{-0.4}$ & $> 1.0$ & $3.0^{+0.5}_{-0.9}$ & $> 11$ & $0.23\pm0.03$ & $5.7^{+7.7}_{-4.5}$ & $0.8^{+0.1}_{-0.2}$ & $0.70\pm0.03$ & 26 & 31.8 & 1.26 (71)\\
		 06/10/15& $2.4^{+0.3}_{-0.2}$ & $2.6^{+0.3}_{-0.2}$ & $1.70\pm0.03$ & 14 & $0.20\pm0.01$ & $3.5^{+3.8}_{-0.8}$ & -- & $4.0^{+0.9}_{-0.7}$ & 2.7 & 6.6 & 1.28 (227)\\
		 08/12/05& $1.7^{+0.6}_{-0.5}$ & $2.3^{+0.5}_{-0.3}$ & $1.7^{+0.05}_{-0.06}$ & 15 & $0.22^{+0.04}_{-0.03}$ & $2.7^{+1.8}_{-1.1}$ & -- & $3.3^{+1.5}_{-0.9}$ & 3.3 & 7.1 & 1.15 (189)\\
		 12/12/16& $2.0\pm0.2$ & $2.6^{+0.2}_{-0.1}$ & $1.73\pm0.02$ & 13 & $0.24^{+0.02}_{-0.01}$ & $2.0^{+0.4}_{-0.3}$ & -- & $2.5^{+0.3}_{-0.3}$ & 2.7 & 6.6 & 1.30 (266)\\
		 12/12/22& $2.0\pm0.1$ & $2.6\pm0.1$ & $1.72^{+0.02}_{-0.03}$ & 13 & $0.25\pm0.01$ & $1.9\pm0.3$ & -- & $2.4^{+0.3}_{-0.2}$ & 2.9 & 6.8 & 1.13 (258)\\
		 13/06/08& $2.2^{+0.8}_{-0.6}$ & $120^{+120}_{-50}$ & $1.83^{+0.08}_{-0.04}$ & $ > 13$ & $0.21^{+0.06}_{-0.04}$ & $2.8^{+3.8}_{-1.5}$ & -- & $3.6^{+2.9}_{-1.2}$ & 3.3 & 6.6 & 1.05 (82)\\
		 14/05/27 (Low)& $4.4\pm0.9$ & $1.6^{+0.3}_{-0.2}$ & $2.19\pm0.1$ & 13 & $0.13^{+0.02}_{-0.01}$ & $28.1^{+21}_{-14}$ & -- & $30^{+23}_{-15}$ & 33.8 & 10.6 & 1.35 (187)\\
		 			& $3.3\pm0.1$ & $> 2.5$ & $3.1^{+0.2}_{-0.6}$ & $>6.2$ & $0.17^{+0.02}_{-0.01}$ & $10.0^{+13.3}_{-7.9}$ & $0.62^{+0.06}_{-0.12}$ & $0.50^{+0.03}_{-0.02}$ & 4.8 & 10.6 & 1.35 (192)\\
		 14/05/27 (Mid)& $4.6\pm1.4$ & $1.6^{+0.3}_{-0.2}$ & $2.1\pm0.1$ & 14 & $0.13^{+0.02}_{-0.01}$ & $35.0^{+41.1}_{-22.9}$ & -- & $37^{+38}_{-21}$ & 51.3 & 15.0 & 1.04 (189)\\
		 			& $3.0\pm0.1$ & $> 2.5$ & $2.7^{+0.6}_{-0.5}$ & $> 5$ & $0.17^{+0.05}_{-0.02}$ & $10.9^{+22}_{-9.8}$ & $0.7^{+0.1}_{-0.2}$ & $0.49^{+0.05}_{-0.04}$ & 7.7 & 15.0 & 0.99 (189)\\
		 14/05/27 (High)& $4.5^{+0.9}_{-1.0}$ & $1.8^{+0.2}_{-0.1}$ & $1.81\pm0.06$ & 15 & $0.13\pm0.01$ & $39.3^{+25.5}_{-18.2}$ & -- & $42^{+25}_{-18}$ & 63.5 & 28.4 & 1.08 (193)\\
		 			& $2.3\pm0.1$ & $2.3^{+1.3}_{-0.3}$ & $2.0^{+0.3}_{-0.2}$ & 11 & $0.18^{+0.05}_{-0.03}$ & $9.0^{+17.8}_{-8.3}$ & $0.7^{+0.3}_{-0.2}$ & $0.54^{+0.05}_{-0.03}$ & 11.1 & 28.3 & 1.01 (197)\\
		 \hline
		 \multicolumn{12}{c}{IC 342 X-1}\\
		 01/01/02 & $1.9^{+0.3}_{-0.7}$ & $1.7^{+1.7}_{-0.3}$ & $< 1.68$ & $>18$ & $0.6^{+0.2}_{-0.4}$ & $0.3^{+3.6}_{-0.1}$ & -- &$0.3^{+3.2}_{-0.1}$ & 1.3 & 3.1 & 0.89 (189)\\
		 04/02/20 & $7.3^{+1.3}_{-1.5}$ & $2.8^{+2.2}_{-0.6}$ & $2.02^{+0.1}_{-0.09}$ & 10 & $0.15\pm0.02$ & $17.9^{17.0}_{-10.4}$ & -- & $19^{+14}_{-9}$ & 23.8 & 7.1 & 1.12 (237)\\
		 		& $2.5$ (fixed) & $2.4^{+1.1}_{-0.5}$ & $1.9^{+0.1}_{-0.2}$ & 12 & $0.5\pm0.1$ & $< 0.3$ & -- & $0.34^{+0.02}_{-0.19}$ & 0.7 & 3.1 & 1.27 (228)\\
		 04/08/17 & $3.4^{+1.3}_{-0.9}$ & $2.5^{+1.5}_{-0.6}$ & $1.7^{+0.1}_{-0.2}$ & 14 & $0.32^{+0.1}_{-0.09}$ & $0.9^{+1.1}_{-0.5}$ & -- & $1.1^{+0.9}_{-0.4}$ & 1.6 & 3.4 & 1.15 (229)\\
		 05/02/10 & $2.3^{+0.8}_{-0.5}$ & $3.3^{+3.4}_{-1.5}$ & $< 2.1$ & $> 9$ & $0.6^{+0.3}_{-0.1}$ & $<0.4$ & -- & $0.38^{+0.03}_{-0.28}$ & 2.4 & 9.5 & 0.93 (197)\\ 
		12/08/11 & $4.2\pm0.8$ & $3.2^{+1.3}_{-0.6}$ & $1.70^{+0.05}_{-0.06}$ & 12 & $0.24^{+0.04}_{-0.03}$ & $2.0^{+1.6}_{-0.9}$ & -- &  $2.4^{+1.3}_{-0.8}$ & 2.5 & 3.54 & 0.92 (275)\\
		 12/08/17 & $4.1\pm1.0$ & $3.0\pm0.2$ & $1.84\pm0.05$ & 11 & $0.22^{+0.07}_{-0.03}$ & $2.2^{+2.6}_{-1.4}$ & -- & $2.9^{+2.0}_{-1.0}$ & 2.54 & 4.0 & 1.07 (291)\\	
		\hline
		\multicolumn{12}{c}{Holmberg IX X-1}\\
		 01/04/23& $1.1^{+0.4}_{-0.2}$ & $1.3\pm0.2$ & $< 1.7$ & $ > 20$ & $0.4\pm0.2$ & $0.7^{+0.5}_{-0.2}$ & -- & $1.0^{+0.4}_{-0.2}$ & 3.3 & 17.4 & 1.18 (72)\\
		 02/04/10& $1.1^{+0.3}_{-0.2}$ & $2.7^{+1.1}_{-0.5}$ & $1.68^{+0.05}_{-0.07}$ & 14 & $0.27^{+0.05}_{-0.04}$ & $1.4^{+0.8}_{-0.4}$ & -- & $2.0^{+0.5}_{-0.3}$ & 2.7 & 10.1 & 1.06 (237)\\
		 02/04/16& $1.5^{+0.5}_{-0.4}$ & $> 3.5$ & $1.78^{+0.05}_{-0.03}$ & $ < 11$ & $0.20^{+0.04}_{-0.03}$ & $2.6^{+2.5}_{-1.3}$ & -- & $4.2^{+1.5}_{-0.8}$ & 3.6 & 11.8 & 1.09 (237)\\
		 04/09/26& $1.6\pm0.2$ & $2.6\pm0.2$ & $1.55\pm0.02$ & 16 & $0.23\pm0.02$ & $2.3^{+0.6}_{-0.5}$ & -- & $2.9^{+0.5}_{-0.4}$ & 3.3 & 9.0 & 1.16 (213)\\
		 11/09/26& $1.5^{+0.4}_{-0.3}$ & $2.6^{+0.6}_{-0.3}$ & $1.84\pm0.03$ & 12 & $0.20^{+0.09}_{-0.05}$ & $<4.7$ & -- & $4.3^{+1.1}_{0.5}$ & 3.8 & 16.7 & 1.05 (237)\\ 
		 11/11/23& $1.1^{+0.2}_{-0.1}$ & $2.6^{+0.4}_{-0.3}$ & $1.82^{+0.04}_{-0.03}$ & 13 & $0.33\pm0.08$ & $< 0.5$ & -- & $1.7^{+0.1}_{-0.2}$ & 4.4 & 20.7 & 1.09 (237)\\ 
		 12/11/12& $0.7\pm0.1$ & $2.1\pm0.1$ & $1.77^{+0.07}_{-0.05}$ & 14 & $0.55\pm0.09$ & $<0.5$ & -- & $0.67^{+0.01}_{-0.03}$ & 5.9 & 22.9 & 1.18 (287)\\	
		 12/11/14& $1.1^{+0.4}_{-0.3}$ & $1.62\pm0.05$ & $1.57\pm0.02$ & 20 & $0.24^{+0.1}_{-0.08}$ & $<2.3$ & -- & $2.9^{+0.7}_{-0.3}$ & 3.6 & 24.5 &  1.22 (276)\\
		 12/11/16& $0.4\pm0.2$ & $2.0^{+0.3}_{-0.1}$ & $1.87^{+0.3}_{-0.1}$ & 18 & $0.77\pm0.2$ & $<0.4$ & -- & $0.38\pm0.01$ & 7.2 & 24.0 & 1.23 (270)\\ 	
		\hline
		\multicolumn{12}{c}{Holmberg II X-1}\\
		 02/04/10 & $0.5\pm0.2$ & $> 3.8$ & $2.59^{+0.05}_{-0.14}$ & $< 18$ & $0.21\pm0.05$ & $<2.4$ & -- & $3.9^{+0.3}_{-0.6}$ & 3.8 & 8.9 & 1.13 (227)\\
		 02/04/16 & $0.4\pm0.2$ & $> 2.0$ & $2.2^{+0.3}_{-0.2}$ & $ < 11$ & $0.22^{+0.03}_{-0.02}$ & $2.3^{+1.2}_{-0.6}$ & -- & $3.5^{+0.8}_{-0.4}$ & 3.8 & 8.7 & 1.20 (202)\\
		 02/09/18 & $0.4^{+0.4}_{-0.3}$ & $0.8^{+1.0}_{-0.3}$ & $< 2.6$ & $>15$ & $0.19^{+0.06}_{-0.04}$ & $2.6^{+2.0}_{-0.9}$ & -- & $3.0^{+1.7}_{-0.8}$ & 3.7 & 1.9 & 0.98 (147)\\
		 04/04/15 & $0.4\pm0.1$ & $3.7^{+5.5}_{-1.0}$ & $2.39^{+0.08}_{-0.07}$ & 7 & $0.21\pm0.03$ & $1.9^{+0.5}_{-0.4}$ & -- & $4.1\pm0.2$ & 4.2 & 9.3 & 1.2 (160)\\
		 10/03/26 & $0.3\pm0.1$ & $1.4^{+0.3}_{-0.2}$ & $2.1\pm0.1$ & 15 & $0.21\pm0.02$ & $2.4^{+0.5}_{-0.4}$ & -- & $2.9^{+0.4}_{-0.3}$ & 2.1 & 3.5 & 1.09 (209)\\ 
		 13/09/09 & $0.7\pm0.3$ & $2.2^{+0.5}_{-0.3}$ & $1.88\pm0.07$ & 13 & $0.19^{+0.03}_{-0.02}$ & $3.8^{+1.9}_{-1.2}$ & -- & $4.7^{+1.6}_{-1.0}$ & 3.7 & 7.0 & 1.05 (246)\\
		 13/09/17 & $0.3^{+0.3}_{-0.2}$ & $2.4^{0.2}_{-0.1}$ & $1.92\pm0.06$ & 12 & $0.20\pm0.03$ & $2.8^{+1.7}_{-1.0}$ & -- & $3.8^{+1.2}_{-0.7}$ & 2.9 & 6.9 &  1.03 (258)\\
		 \hline
		\end{tabular}
		\begin{flushleft}
			{\footnotesize
			Notes: $^a$Date of the observations.
			$^b$Intrinsic column density of equivalent Hydrogen in units of $10^{21}$~cm$^{-2}$.
			$^c$Optical depth of the coronal electron cloud. See text for the statistical uncertainty.
			$^d$Inner-disk temperature or that at the radius where the corona is truncated (see text).
			$^e$Apparent inner-disk radius of the un-scattered accretion disk component in units of 1000~km.
			$^f$Inner-disk radius of the overall disk component in units of 1000~km. Calculated from equation \ref{eq:rtot}.
			$^g$Bolometric luminosity of the accretion disk component in units of $10^{39}$~erg~sec$^{-1}$.
			$^h$Absorbed $0.3\--10$~keV band luminosity in units of $10^{39}$~erg~sec$^{-1}$.
			$^i$Apparent inner-disk radius of the outer-disk region in units of 1000~km.
			$^j$Inner-disk temperature of the model representing the inner part of the disk in the alternative model (see text).
			}
		\end{flushleft}
		\end{table*}

		\begin{table*}
		\contcaption{}
		\label{tab:fitparam2}
		\begin{tabular}{cccccccccccc}
		\hline
		\hline
		Date$^{\rm{a}}$ & $N_{\rm{H}}$$^{\rm{b}}$ & $T_{\rm{e}}$ & $\Gamma$ & $\tau$$^{\rm{c}}$ & $T_{\rm{in}}$ or $T_{\rm{in1}}$$^{\rm{d}}$ & $R_{\rm{raw}}$$^{\rm{e}}$ or & $T_{\rm{in2}}$$^{\rm{j}}$ & $R_{\rm{tot}}$$^{\rm{f}}$ & $L_{\rm{disk}}$$^{\rm{g}}$ & $L_{\rm{X}}$$^{\rm{h}}$ & $\chi^{2}/\nu$ ($\nu$)\\
		YY/MM/DD& & (keV) & & & (keV) & $R_{\rm{raw1}}$$^{\rm{i}}$ & (keV) & & & & \\
		\hline\hline
		 \multicolumn{12}{c}{M33 X-8}\\
		 00/08/04 & $0.3^{+0.2}_{-0.1}$ & $1.19^{+0.2}_{-0.08}$ & $1.8^{+0.2}_{-0.1}$ & 19 & $0.4\pm0.1$ & $< 0.2$ & -- & $0.40^{+0.02}_{-0.22}$ & 0.5 & 1.7 & 1.21 (204)\\
		 00/08/07 & $0.2^{+0.2}_{-0.1}$ & $1.6^{+1.6}_{-0.7}$ & $<3.2$ & $>8$ & $0.53^{+0.3}_{-0.07}$ & $< 0.3$ & -- & $0.22^{+0.01}_{-0.15}$ & 0.4 & 1.6 & 1.12 (172)\\
		 03/02/12 & $0.25^{+0.09}_{-0.08}$ & $2.2^{+1.1}_{-0.4}$ & $2.1\pm0.1$ & 11 & $0.41^{+0.08}_{-0.04}$ & $<0.1$ & -- & $0.34^{+0.01}_{-0.04}$ & 0.3 & 1.4 & 1.10 (213)\\
		 10/01/11 & $0.12\pm0.08$ & $1.9^{+0.2}_{-0.1}$ & $2.06^{+0.08}_{0.06}$ & 12 & $0.5\pm0.05$ & $< 0.06$ & -- &$0.230^{+0.003}_{-0.011}$ & 0.3 & 1.5 & 1.27 (136)\\	
		\hline
		\multicolumn{12}{c}{M83 ULX-1}\\
		 13/08/07 & $< 0.4$ & $1.35^{+0.1}_{-0.09}$ & $1.58^{+0.05}_{-0.1}$ & 22 & $0.28^{+0.16}_{-0.03}$ & $< 0.6$ & -- &$0.75^{+0.03}_{-0.08}$ & 0.5 & 3.1 & 1.08 (222)\\
		 14/01/11 & $0.8^{+0.7}_{-0.5}$ & $0.9^{+0.3}_{-0.2}$ & $1.6\pm0.3$ & 28 & $0.21^{+0.09}_{-0.05}$ & $0.8^{+1.2}_{-0.4}$ & -- & $1.0^{+1.0}_{-0.4}$ & 0.3 & 1.1 & 0.82 (171)\\
		 14/07/06 & $ < 8.2$ & $1.3\pm0.1$ & $1.6^{+0.07}_{-0.2}$ & 21 & $0.28^{+0.2}_{-0.05}$ & $<  0.3$ & -- & $0.67^{0.02}_{-0.15}$ & 0.4 & 2.3 & 1.03 (227)\\
		 15/02/02 & $1.1^{+1.1}_{-0.8}$ & $0.9^{+0.3}_{-0.3}$ & $1.7^{+0.2}_{-0.4}$ & 23 & $0.17^{+0.1}_{-0.05}$ & $< 0.6$ & -- & $2.0^{+3.4}_{-1.2}$ & 0.5 & 0.8 & 0.81 (150)\\ 
		\hline
		\multicolumn{12}{c}{M83 ULX-2}\\
		  03/01/27 & $0.4^{+0.4}_{-0.3}$ & $0.71^{+0.2}_{-0.07}$ & $< 1.7$ & $> 27$ & $0.29\pm0.06$ & $0.7^{+0.3}_{-0.2}$ & -- & $0.7^{+0.3}_{-0.2}$ & 0.4 & 0.8 & 1.26 (143)\\
		 13/08/07 & $0.2^{+0.2}_{-0.1}$ & $0.76^{+0.2}_{-0.05}$ & $< 2.1$ & $> 19$ & $0.40^{+0.05}_{-0.1}$ & $0.4\pm0.1$ & -- & $0.4^{+1.1}_{-0.1}$ & 0.5 & 1.2 & 1.09 (195)\\
		 14/01/11 & $0.1^{+0.2}_{-0.1}$ & $0.9^{+0.2}_{-0.1}$ & $< 2.1$ & $>17$ & $0.4\pm0.1$ & $< 0.5$ & -- & $0.5^{+0.1}_{-0.3}$ & 0.8 & 2.5 & 1.12 (217)\\
		 14/07/06 & $0.5\pm0.3$ & $0.73^{+0.3}_{-0.08}$ & $< 2.1$ & $20.1$ & $0.3\pm0.1$ & $0.5^{+0.3}_{-0.2}$ & -- & $0.6^{+0.3}_{-0.2}$ & 0.4 & 1.2 & 1.15 (177)\\ 
		\hline
		\multicolumn{12}{c}{NGC 4190 X-1}\\
		  10/06/06 & $0.5\pm0.4$ & $0.83^{+0.15}_{-0.04}$ & $< 2.2$ & $> 18 $ & $0.41^{+0.08}_{-0.2}$ & $0.5^{+0.6}_{-0.1}$ & -- & $0.5^{+0.6}_{-0.1}$ & 0.9 & 3.6 & 1.15 (202)\\
		  10/06/08 & $0.9^{+0.6}_{-0.4}$ & $1.08^{+0.09}_{-0.1}$ & $1.55^{+0.09}_{-0.12}$ & 25 & $0.2^{+0.2}_{-0.1}$ & $<1.7$ & -- & $1.4^{+0.9}_{-0.4}$ & 0.4 & 4.5 & 1.19 (197)\\
		  10/11/25 & $0.6\pm0.2$ & $1.6^{0.2}_{-0.3}$ & $<1.7$ & $ > 18$  & $0.42^{+0.5}_{-0.1}$ & $<0.5$ & -- & $0.63^{+0.03}_{-0.50}$ & 1.2 & 8.5 & 1.08 (237)\\
		\hline
		\multicolumn{12}{c}{NGC 1313 X-2}\\
		 12/12/16 & $1.7^{+0.3}_{-0.2}$ & $1.17\pm0.05$ & $1.69^{+0.05}_{-0.08}$ & 21 & $0.28^{+0.09}_{-0.06}$ &  $0.6^{+0.5}_{-0.2}$ & -- & $1.3^{+0.2}_{-0.1}$ & 1.3 & 4.8 & 1.16 (239)\\
		 12/12/22 & $1.9^{+0.5}_{-0.3}$ & $1.07\pm0.09$ & $2.02^{+0.08}_{-0.13}$ & 17 & $0.21^{+0.08}_{-0.06}$ & $<2.4$ & -- & $2.0^{+0.7}_{-0.5}$  & 1.0 & 2.2 & 1.14 (229)\\
		\hline
		\end{tabular}
		\begin{flushleft}
			{\footnotesize
			Notes: $^a$Date of the observations.
			$^b$Intrinsic column density of equivalent Hydrogen in units of $10^{21}$~cm$^{-2}$.
			$^c$Optical depth of the coronal electron cloud. See text for the statistical uncertainty.
			$^d$Inner-disk temperature or that at the radius where the corona is truncated (see text).
			$^e$Apparent inner-disk radius of the un-scattered accretion disk component in units of 1000~km.
			$^f$Inner-disk radius of the overall disk component in units of 1000~km. Calculated from equation \ref{eq:rtot}.
			$^g$Bolometric luminosity of the accretion disk component in units of $10^{39}$~erg~sec$^{-1}$.
			$^h$Absorbed $0.3\--10$~keV band luminosity in units of $10^{39}$~erg~sec$^{-1}$.
			$^i$Apparent inner-disk radius of the outer-disk region in units of 1000~km.
			$^j$Inner-disk temperature of the model representing the inner part of the disk in the alternative model (see text).
			}
		\end{flushleft}
		\end{table*}

\subsection{Search for Fe-K line features}
\begin{figure*}
	\includegraphics[width=\textwidth]{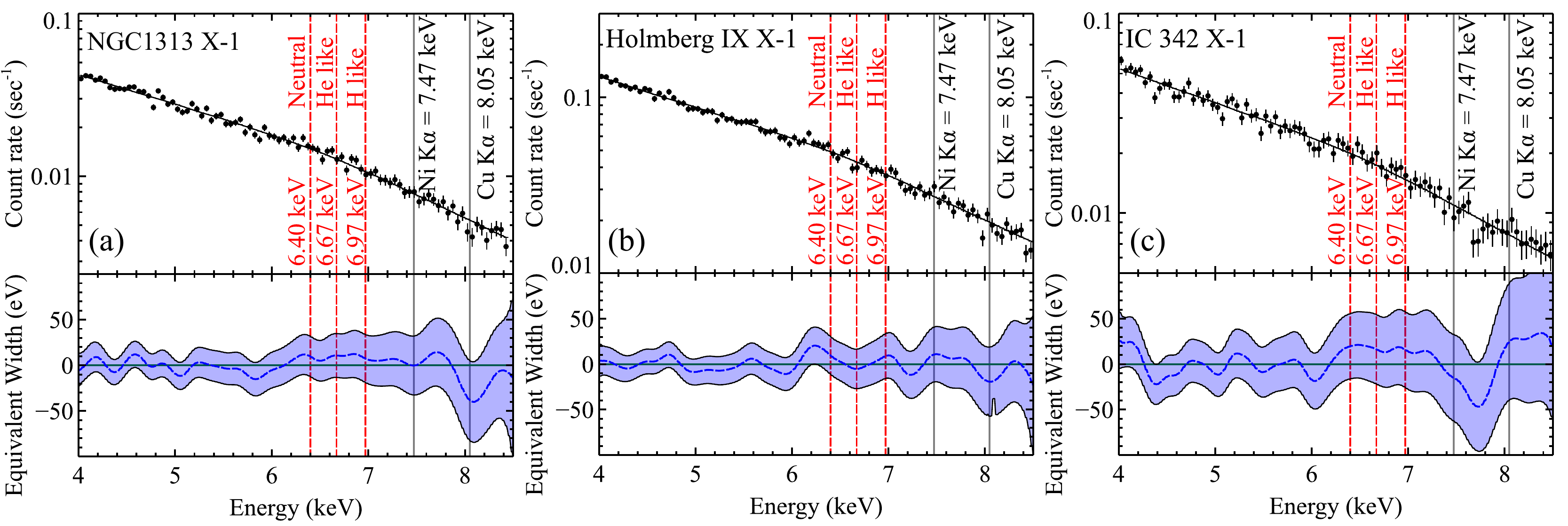}
	\caption{Stacked PN spectra of NGC 1313 X-1 (panel a), Holmberg IX X-1 (panel b), and IC 342 X-1 (panel c), fitted with a cutoff power law model (top panels). The bottom panels indicate the allowed range of the line equivalent width, where the blue dashed line and the light blue band represent the best fit value and the $99\%$ confidence range, respectively. Positive and negative equivalent widths specify emission and absorption lines, respectively.}
	\label{fig:ew_others}
\end{figure*}
As described in section 3.5, the MCD+THC model successfully explained all the sample spectra, and none of them exhibited noticeable emission/absorption line features (except those of $<1$~keV; section 3.2.1), including in particular Fe-K lines which are widely seen in X-ray spectra of various classes of accreting objects. 
Although the ULX spectra in our sample are thus generally featureless, each individual spectrum could still be insufficient in statistics to derive meaningful constraints on the Fe-K line features. 
Therefore, we have decided to co-add (stack) the separate spectra of each source into a single one, and search it for weak Fe-K lines under improved statistics.

Since the analysis requires high photon flux around the Fe-K edge energy (7.11~keV), we selected for this purpose three sources, NGC 1313 X-1, IC 342 X-1, and Holmberg IX X-1, which exhibited hard spectra with $\Gamma \sim 1.7$ and the highest X-ray flux ($> 10^{-11}$ erg sec$^{-1}$ cm$^{-2}$) among our sample. 
To avoid the obvious complication of adding up the data from different instruments, we limit the stacking analysis to the \textit{XMM-Newton} data which are rather abundant, and utilize only those of EPIC PN which has a larger effective area than EPIC MOS at $\sim7$~keV. 

The top panels of figure \ref{fig:ew_others} show the stacked PN spectra of the three sources, NGC 1313 X-1, Holmberg IX X-1, and IC 342 X-1, with a total exposure of 394~ks (14 spectra), 132~ks (11 spectra), and 113~ks (6 spectra), respectively. Since our interest is in Fe-K lines, we restricted the analyzing energy band to $4.0\--8.5$~keV, which is still wide enough to allow blue/red shifts up to $20\%$ of the speed of light.

To quantify the continuum in this energy band, we fitted each stacked spectrum with a cutoff-PL model, as represented by the solid line superposed on the spectrum. 
This has already given a good fit with $\chi^{2}/\nu = 133.0/116$, $\chi^{2}/\nu = 80.34/86$, and $\chi^{2}/\nu = 87.7/96$, in NGC 1313 X-1, Holmberg IX X-1, and IC 342 X-1, respectively, where $\nu$ is the degree of freedom. 
In the fit residuals, we do not see noticeable emission/absorption features, except a possible dip at  $\sim8$~keV in NGC 1313 X-1, which is likely to have arisen from an over subtraction of the instrumental Cu-K line in the background spectrum. 

To derive upper limits on the emission/absorption features, to the model we added a Gaussian with a width of $10$~eV at various energies, allowed its normalization to take either positive or negative values, and scanned the energy with a step of $10$~eV to calculate $99\%$-confidence upper limits on the allowed line equivalent width (EW). 
The results are given in light-blue strips in the bottom panels of figure \ref{fig:ew_others}. 
Thus, the stacked spectra have provided rather stringent limits on emission or absorption line features. 
Especially, around the energy where the Fe-K features without blue/red shifts are expected (red dashed lines in figure \ref{fig:ew_others}), we obtained $99\%$ upper limits on the EW as $30$~eV in NGC 1313 X-1, 35~eV in Holmberg IX X-1, and 55 eV in IC 342 X-1. 

\section{Discussion}
\subsection{Summary of the Results}
Utilizing \textit{Suzaku}, \textit{XMM-Newton}, and \textit{NuSTAR}, we studied 56 spectra of 9 representative ULXs, and obtained the following results.
\begin{enumerate}[leftmargin=10pt]

\item Of the 56 spectra, 3 can be classified into the SPL state, 24 into the HPL state, and the remaining 19 into the MCD state. When a source resided in multiple states, its luminosity was always in the order of SPL $<$ HPL $<$ MCD.

\item In any spectral state, the spectra in the present sample were all successfully reproduced by a single continuum model (the MCD+THC model), requiring no additional local feature components.

\item Regardless of the spectral states, the 9 sources are all inferred to harbor cool ($T_{\rm{e}}\sim1.5\--3$~keV) and optically-thick ($\tau > 10$) corona, in agreement with the results of some previous works (e.g., \citealt{Miyawaki2009}).

\item The spectral transition from the SPL to HPL states is characterized by an increase in the $T_{\rm e}/T_{\rm in}$ ratio, whereas that from the HPL to MCD states by an increase in $T_{\rm in}$ and a marked decrease in the directly-visible-disk fraction.

\item All sources showed stable and low absorption as $N_{\rm{H}}\sim10^{21}$~cm$^{-2}$, where the Galactic line-of-sight contribution is separately modeled and removed.

\item In three sources with high signal statistics, Holmberg IX X-1, IC 342 X-1, and NGC 1313 X-1, stringent upper limits of $30\--55$~eV in EW were obtained for narrow Fe K-shell emission/absorption lines.

\end{enumerate}

\subsection{Spectral State Transitions and State Indicators}
In section 3, many of the sources in our sample showed multiple spectral states from SPL, HPL, and MCD, with the luminosity increasing in this order [section 4.1 item (i)]. 
At the same time, the 56 spectra were all fitted successfully with a single unified model, namely, the MCD +THC model [section 4.1 item (ii)]. 
Then, as already suggested by section 4.1 item (iv), we must be able to utilize the best-fit model parameters to quantitatively specify the spectral states, instead of just looking by eyes at their spectral shapes.

To characterize the spectral shapes of ULXs, we introduce three quantities, each derived from the obtained model parameters. 
They are the temperature ratio $Q\equiv T_{\rm{e}}/T_{\rm{in}}$, the coronal covering fraction
\begin{equation}
F\equiv 1- (R_{\rm{raw}}/R_{\rm{tot}})^{2}~,
\label{eq:f}
\end{equation}
and the Compton y-parameter. 
Here, $Q$ represents the balance point between heating and cooling of the coronal electrons \citep{Zhang2016}, and $F$ describes the fractional area of the accretion disk covered by the corona \citep{Kobayashi2016}. 
Although the radii $R_{\rm{raw}}$ and $R_{\rm {thc}}$ are affected by distance uncertainties and the unknown inclination angle of the accretion disk, these should cancel out by taking ratios in the calculation of $F$. Hence, $F$ is a parameter which is directly comparable among different sources. 

The third parameter, $y$, needs some caution.
If the coronal electron temperature $T_{\rm{e}}$ is much higher than that of the seed photons (i.e., $T_{\rm{e}} \gg T_{\rm{in}}$ in the present case), the y-parameter is generally approximated, with the help of equation \ref{eq:tau}, as $y\sim4kT_{\rm{e}}\tau(1+\tau/3)/m_{\rm{e}}c^{2}=4/[(0.5+\Gamma)^2-2.5]$ in an optically thick condition, where $k$ and $m_{\rm{e}}$ are the Boltzmann constant and the electron mass, respectively. 
However, the temperature of $T_{\rm{e}}\sim 2$~keV found in our sample is rather too low to be regarded as $T_{\rm{e}} \gg T_{\rm{in}}$, because it is at most only 10 times higher than the seed photon temperature, $T_{\rm{in}} \sim 0.2$~keV. 
To take this closeness into account, we hereafter modify the definition as 
\begin{equation}
y\equiv \frac{4k(T_{\rm e}-T_{\rm in})}{m_{\rm e}c^{2}}\tau(1+\tau/3)=\frac{4k}{[(0.5+\Gamma)^2-2.5]}\left(\frac{T_{\rm e}-T_{\rm in}}{T_{\rm e}}\right)~,
\label{eq:y_rev}
\end{equation}
following \citet{Zhang2016} who studied similarly cool and thick coronae.

\begin{figure}
\begin{center}
\includegraphics[width=\columnwidth]{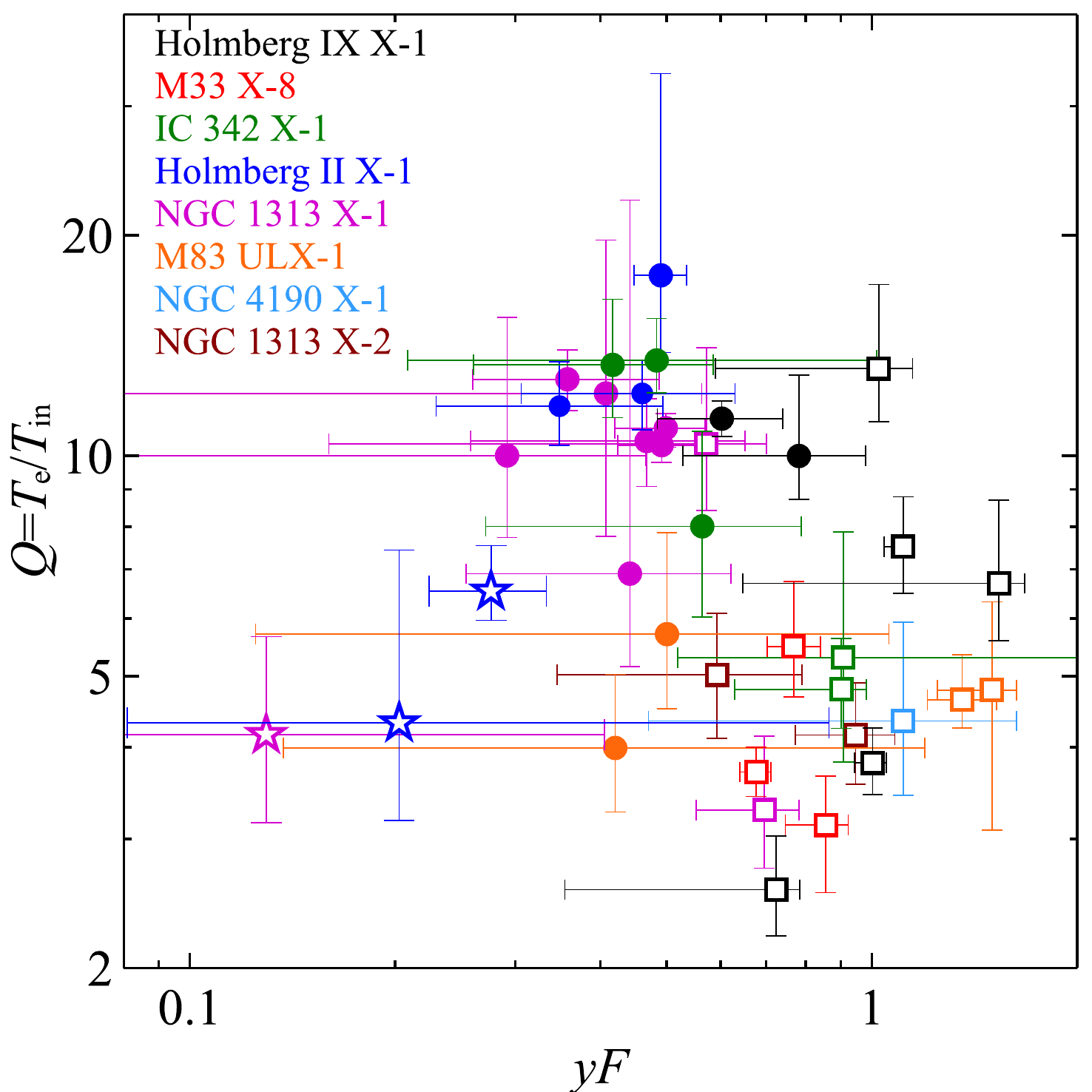}
\caption{A scatter plot between $yF$ and $Q$ for the present spectra. The open stars, the filled circles, and the open squares represent the data in the SPL, the HPL, and the MCD states, respectively. Colors specify the different objects. The error bars correspond to 68\% confidence level.}
\label{fig:fy_vs_q}
\end{center}
 \end{figure}
In terms of these physical/geometrical parameters, let us try classifying the three spectral states in an objective way. 
Since there are three states, we may need at least two parameters, of which one would be $Q$ itself because of section 4.1 item (iv).
As the other, the product $y\times F$ is expected to be promising: $F$ describes how much fraction of the disk photons get Comptonized, whereas $y$ (equation \ref{eq:y_rev}) specifies the relative energy gain of each Comptonized photon, so their product would provide a good measure of the degree of Comptonization of all photons emerging from each system.

In figure \ref{fig:fy_vs_q}, we present a scatter plot between $y\times F$ and $Q$ for 39 spectra. 
For clarity, we omitted 17 data points with poor statistics, in which either $T_{\rm{e}}$ or $\Gamma$ were unconstrained. 
The spectral states, already identified by their overall spectral appearance, are distinguished by different symbols.
Thus, the plot successfully accomodates the SPL and MCD data points at $Q \le 10$ due to the closeness between $T_{\rm{in}}$ and $T_{\rm{e}}$, and those in the HPL state at $Q > 10$ due to the large difference between the two temperatures.
At the same time, we observe that the SPL, HPL, and MCD data points generally appear on figure \ref{fig:fy_vs_q} as the increasing order of the $yF$ product. 
This means that the overall Comptonization effect is highest in the MCD state, and lowest in the SPL state.
As a results of these two tendencies, the data points draw a convex locus on the $(yF,~Q)$ plane. 
Thus, the $yF$ vs. $Q$ plot has successfully distinguished the three characteristic spectral states in a quantitative way, even though the errors are rather large.

\subsection{Discrimination of NS ULX Candidates}
During the revision of this paper, weak coherent pulsations with a period of $\sim1.5$ s were discovered in some of the latest X-ray observations of NGC 1313 X-2 \citep{Sathyaprakash2019}. Although the statistical significance of the detection is still marginal ($3\--4 \sigma$), the result strongly suggests that the source, long believed to be an accreting BH, actually harbors an NS as the mass accretor. As we have seen so far, the spectra of NGC 1313 X-2 are apparently similar to those of the other ULXs in the MCD state. Hence, there arises a possibility that our sample could still be contaminated by NS ULXs, besides NGC 1313 X-2. To minimize this risk, we need to look for some additional ways to distinguish potential NSs from the rest.

\begin{figure}
\begin{center}
\includegraphics[width=\columnwidth]{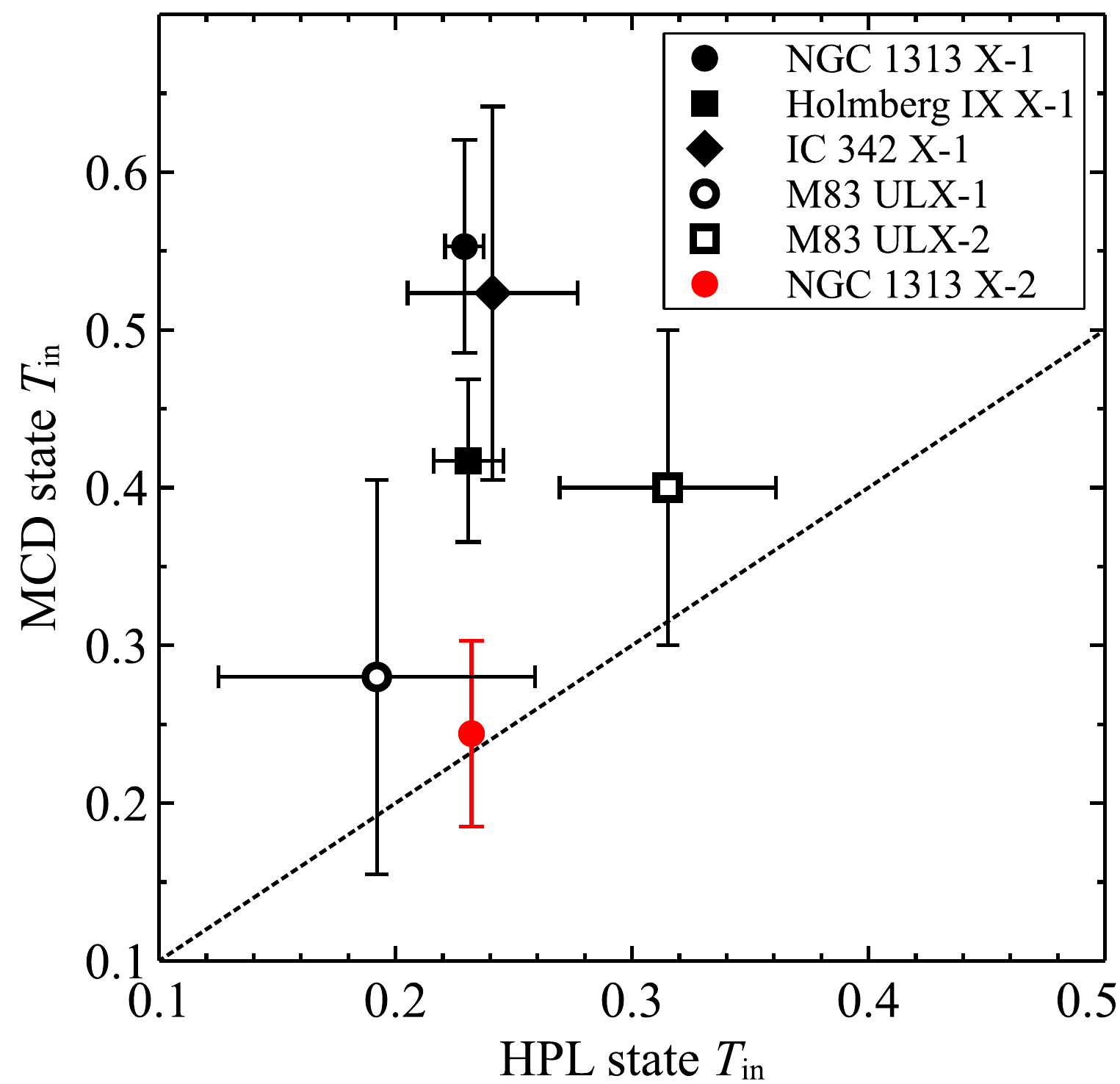}
\end{center}
\caption{Our sample objects, plotted on the plane of the error weighted mean $T_{\rm in}$ in the HPL state (abscissa) and that in the MCD state (ordinate). The dashed line indicates that the two temperatures are equal. The plot excludes those sources which did not make transitions between the HPL and the MCD states in our observations. See text for the treatment of NGC 1313 X-2 (shown in red). }
\label{fig:thpl_tmcd}
\end{figure}
As an attempt, in figure \ref{fig:thpl_tmcd} we plot our sample objects on the plane of the error-weighted mean $T_{\rm in}$ in the HPL states and that in the MCD state. Since the present sample does not include any spectra of NGC 1313 X-2 in the HPL state, we took its HPL $T_{\rm in}$ values from \citet{Pintore2012}, in which several HPL state spectra of NGC 1313 X-2 were analyzed with the same type of spectral modeling as the present work. Thus, most of the sources show a significantly higher $T_{\rm in}$ in the MCD state than in the HPL state, whereas NGC 1313 X-2 exhibits insignificant change in $T_{\rm in}$ between the two states. Although the physical origin of this difference is unknown, the result may allow us to discriminate NGC 1313 X-2 (possibly M83 ULX-1 as well) from the other objects in the sample. Since no other NS ULXs are reported to have made transitions between the HPL and MCD states, we are not sure at present whether this discrimination is general or not, but we employ figure \ref{fig:thpl_tmcd} as a criterion to remove NGC 1313 X-2 from our subsequent discussion.

\subsection{State Transition Luminosities}
\begin{figure*}
  \begin{center}
   \includegraphics[width=\textwidth]{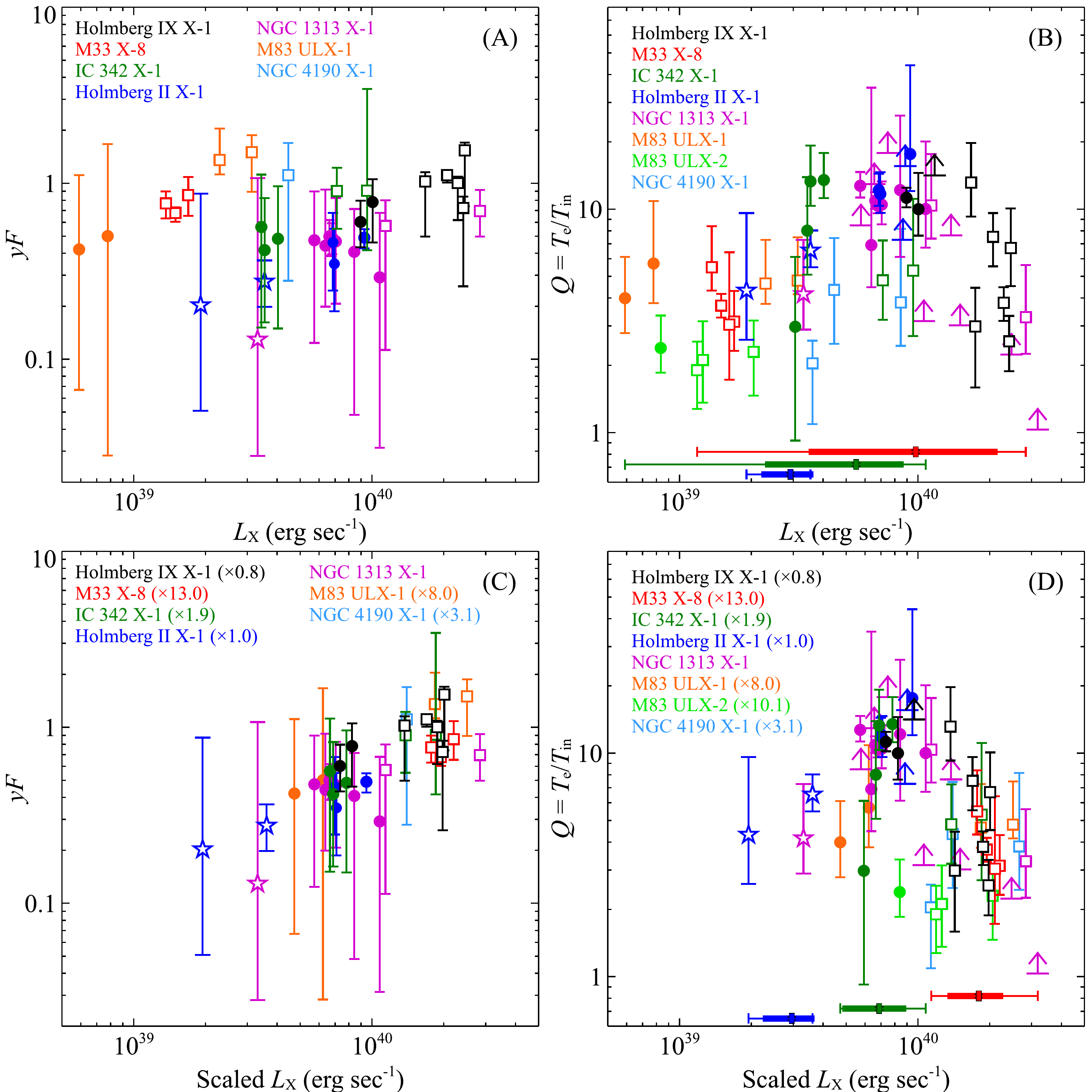}
  \end{center}
  \caption{The parameters $Q$ (panel A) and $yF$ (panel B) of the present spectra, shown against the $0.3\--20.0$~keV luminosity $L_{\rm{X}}$. The plotting symbols and the colors are the same as in figure \ref{fig:fy_vs_q}. (C) and (D) are the same panels as (A) and (B), respectively, but the luminosity of each source is scaled by a certain factor to match that of NGC 1313 X-1. The scaling factors are shown in the legends. The horizontal bars at the bottom of panels (B) and (D) represent the luminosity distributions of the individual spectral states (blue for SPL, green for HPL, red for MCD). The central markers, thin horizontal lines, and thick bars show their mean values, full distribution ranges, and the standard deviations, respectively.}
  \label{fig:l_vs_qandyf}
\end{figure*}

As described in section 1, one of our interests is how the spectral states are distributed against the source luminosity. 
Since we have shown above that the two new parameters $Q$ and $yF$ are good indicators of their spectral states, we present in figure \ref{fig:l_vs_qandyf} their behavior as a function of the $0.3\--20$~keV luminosity $L_{\rm{X}}$.

In contrast to the relatively well-organized locus seen in figure \ref{fig:fy_vs_q}, the $L_{\rm{X}}$ vs. $Q$ and $L_{\rm{ X}}$ vs. $yF$ plots, figure \ref{fig:l_vs_qandyf} (A) and figure \ref{fig:l_vs_qandyf} (B) respectively, both suffer significant scatter of the data points.
In fact, almost no correlation is seen in figure \ref{fig:l_vs_qandyf} (A), and the distribution ranges of the three states, represented by three colored horizontal bars in figure \ref{fig:l_vs_qandyf} (B), heavily overlaps with one another.
We consider that this is because the individual sources reside in a particular spectral state (i.e., similar values in $Q$ or $yF$) at rather different luminosities.
In fact, M33 X-8 exhibit a series of MCD-state spectra (open squares) at $L_{\rm{X}}\sim1.5\times 10^{39}$ erg ~sec$^{-1}$, whereas the sources such as Holmberg IX X-1 and NGC 1313 X-1 at $L_{\rm{X}}\sim3\times 10^{40}$~erg~sec$^{-1}$. 
Then, the data scatter in these panels would diminish if we scale $L_{\rm{X}}$ by appropriate factors that are specific to the individual sources.

As presented in panels (C) and (D) of figure \ref{fig:l_vs_qandyf}, we hence scaled $L_{\rm {X}}$ of 7 sources (except NGC 1313 X-1), by a factor which brings the behavior of each source to line up with that of NGC 1313 X-1. We scanned the scaling factors of the individual sources over appropriate ranges, and searched for the optimum values that minimize the overall data scatter in $L_{\rm X}$ in the following way. First, in the $i$-th spectral state ($i=1,2,3$), the average $L_{\rm X}$ of NGC 1313 X-1 was chosen as the fiducial value. Then, relative to that value, the standard deviation $\sigma_i$ of the scaled luminosity of the other spectra in the same state was calculated. Finally, the overall scatter was calculated as $\sigma = \sum_{i=1}^{3} \sigma_i$. We have employed the standard deviation rather than chi-square, because errors associated with individual $L_{\rm X}$ determinations are difficult to estimate and are considered to be relatively similar among the sample spectra. The obtained optimum scaling factors are given in the legends of these panels.
As a result, the data scatter in the $L_{\rm{X}}$ vs. $Q$ and $L_{\rm{X}}$ vs. $yF$ plots have both decreased significantly.
At the same time, a good correlation has appeared in figure \ref{fig:l_vs_qandyf} (C) between $yF$ and the scaled $L_{\rm X}$. In figure \ref{fig:l_vs_qandyf} (D), the three states have become re-distinguishable in terms of $Q$ and $yF$ just as in figure \ref{fig:fy_vs_q}.
These results indicate that the luminosity renormalization is physically meaningful, and the state-emerging luminosity systematically differs by at least a factor of $\sim 16$ among the 8 sources.

The canonical BH/NS binary systems also show several spectral states as a function of their luminosity, and the behavior is almost uniquely determined by the Eddington ratio $\eta$.
For example, the standard High/Soft State are observed typically at $\eta=0.1\--0.3$, and so-called Very High State of BHBs emerges at $\eta \sim 0.3$ (e.g., \citealt{Kubota2004}; \citealt{Tamura2012}). 
If we assume that this general property widely seen in accreting systems is also shared by ULXs, the luminosity re-normalizing factor which we have employed can be regarded, most naturally, as the $(L_{\rm edd})_{\rm{\circ}}/L_{\rm{edd}}$ ratio of the each source, where $(L_{\rm edd})_{\rm{\circ}}$ refers to the value of NGC 1313 X-1. In other words, the scaled luminosity can be identified with $\eta$, except a common proportionality factor. This implies that $L_{\rm edd}$ of our sample sources also scatters by the same factor ($\sim 16$). 

Since $L_{\rm edd}$ is proportional to the mass of the compact object, the difference in $L_{\rm edd}$ directly suggests that the mass of our sample also scatters over a similar range. 
For example, we may assume that the lowest mass (M33 X-8) of the present ULXs to be $\sim 10\ M_{\rm \odot}$, which is typical of a stellar mass BH. 
Then, the highest mass in our sample (Holmberg IX X-1) should reach $\sim160 \ M_{\rm \odot }$, which is in the intermediate mass regime.
If so, the 8 ULXs (excluding NS ULX NGC 1313 X-2 which is very likely to be an NS-ULX) are implied to have a broad mass distribution from $\sim 10\ M_{\rm \odot}$ to $\sim 160 \ M_{\rm \odot}$, in qualitative agreement with the latest results from the Gravitational Wave experiments.
Thus, unlike the ordinary BH/NS binaries, but somewhat similar to Active Galactic Nuclei, these typical ULXs are suggested to have a wide mass range, of which the highest end will exceed the generally accepted masses of stellar BHs.
The values of $R_{\rm{in}}$ (table \ref{tab:fitparam}), typically of the order of $\sim10^{3}$~km, are consistent with this interpretation, even though these estimates are subject to the large statistical (and possibly systematic as well) uncertainties.
This argument assumes only a general property which is widely seen in the accreting objects, and do not rely on any particular accretion models. 
In particular, the argument does not rely on absolute values of $\eta$, and hence should not depend on whether ULXs are super-Eddington or sub-Eddington sources.

\subsection{Paucity of local materials}
\subsubsection{Absorption line features}
As revealed in section 3, the present stacked ULX spectra are devoid of spectral evidence for significant materials around the sources. To consider implications of these results, we first compare the strength of the Fe K$\alpha$ absorption lines of the present sample with those obtained in other accreting objects.
 
Blue-shifted Fe K$\alpha$ absorption lines with high ionization are occasionally seen in spectra of some BH binaries in the High/Soft state (e.g., \citealt{Miller2006}) and active galactic nuclei \citep{Tombesi2010}. This particular feature is evidently telling us that some ``winds'' are often launched into our line of sight, presumably from the accretion disks. Since ULXs are as powerful accretors as these objects, it is natural to expect these winds to be also present in ULXs. Hence, many attempts were made to search their spectra for these absorption or emission line features. In particular, stacking analysis similar to ours have been made with other data of bright ULXs (e.g., \citealt{Walton2013}). Despite these efforts including ours, none of the ULX analyzed showed noticeable signs of the Fe K$\alpha$ absorption lines, except for one particular observation of NGC 1313 X-1 (velocity of $\sim0.2c$ and EW $-61\pm24$ eV; \citealt{Walton2016}). Furthermore, upper limits on the EW of absorption lines, obtained from the other observations of ULXs, are typically several tens eV like in the present case, which is comparable to or more stringent than the actual detections from BHBs in the High/Soft state ($20\--70$~eV; e.g., \citealt{Ponti2012, Yamada2013}). Gathering these results together, the disk winds in ULXs are likely to be generally rather weak or transient.

The detection of the blue-shifted highly-ionized lines from NGC 1313 X-1 could be taken as a good evidence for ULXs being supercritical accretion states.
In fact, some numerical simulations show that massive disk winds are launched in such an accretion regime (e.g., \citealt{Kawashima2012}).
However, such disk winds are also confirmed in BH binaries that clearly reside in sub-critical regimes (e.g., \citealt{Miller2012, Ponti2012}).
Furthermore, some sources that are likely to be accreting at supercritical rate are known to yield significant Fe K absorption-edge features as well (e.g., \citealt{Hagino2016, Shidatsu2016}), which are not observed from ULXs.
Thus, detections of blue-shifted absorption lines would not be regarded as unique evidence of supercritical accretion flows.

\subsubsection{Emission line features and low energy absorption}
\begin{figure}
	\begin{center}
	\includegraphics[width=\columnwidth]{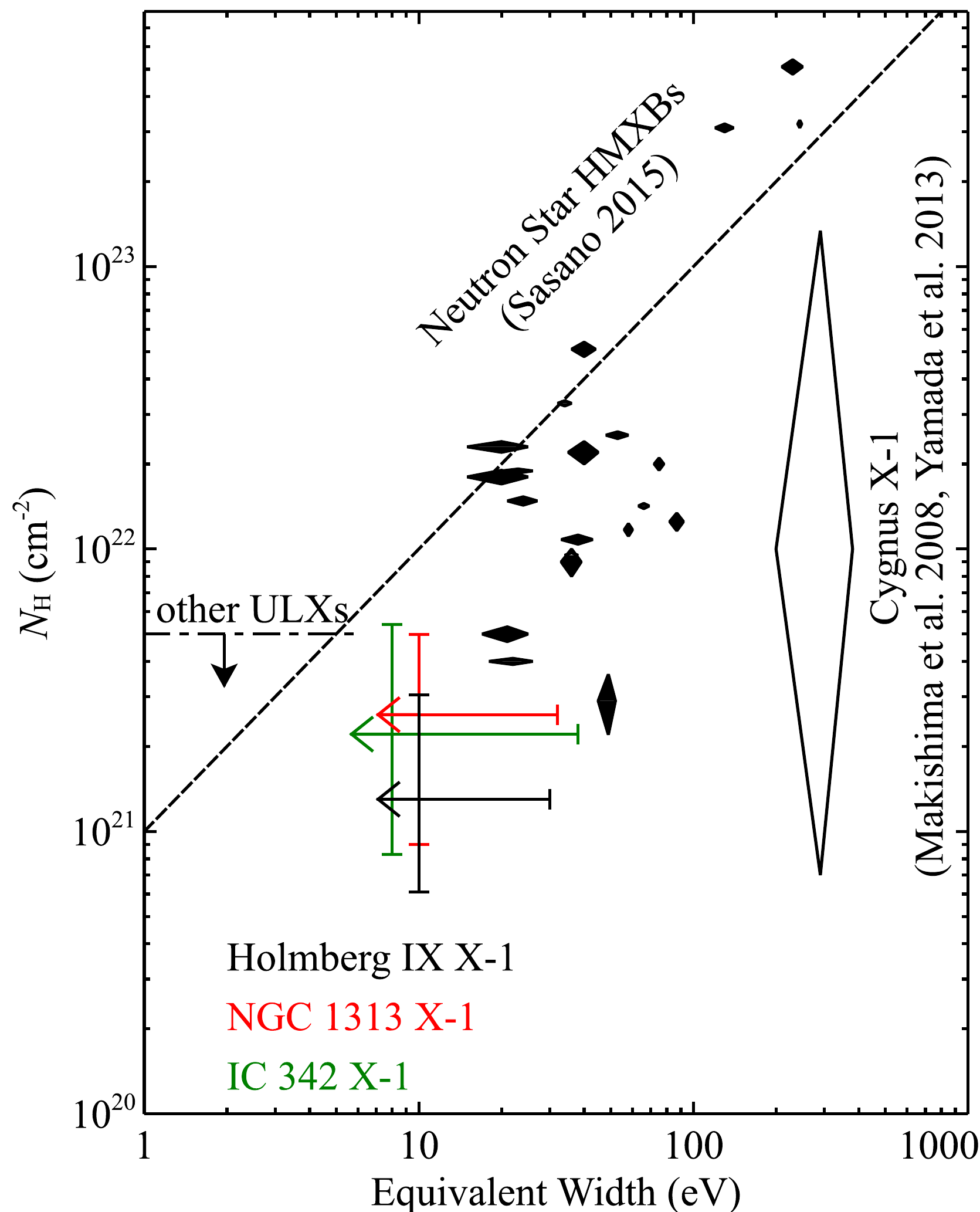}
	\caption{A scatter plot between the Fe K$\alpha$ line equivalent width and the line-of-sight absorption column density $N_{\rm{H}}$. The Galactic contributions in $N_{\rm{H}}$ of ULXs were removed, while those of the other classes of objects are not. The dot-dashed line indicates the maximum value obtained from the present ULX sample. The dashed line represents predictions for isotropic and spherical gas clouds with a solar abundance \citep{Inoue1985}, neglecting self-absorption of the Fe-K lines which become significant at $N_{\rm{H}} \gtrsim 10^{23}$~cm$^{-2}$}
	\label{fig:n_vs_ew}
	\end{center}
\end{figure}

Next, we examine the absorbing column densities $N_{\rm H}$ and the Fe-K line EWs, measured in the present study, utilizing the EW vs $N_{\rm H}$ diagram in figure \ref{fig:n_vs_ew}. While the line EW represents the total amount of matter surrounding the source, $N_{\rm H}$ gives the amount of matter in the line of sight. If the matter surrounds the central source spherically and uniformly, these two quantities should be proportional to each other, as shown by the dotted line in figure \ref{fig:n_vs_ew} \citep{Inoue1985}. Since the matter distribution is often highly anisotropic, considering both the line EW and $N_{\rm H}$ provides a good tool of diagnostics.

On the plane of EW vs $N_{\rm H}$, figure \ref{fig:n_vs_ew} compares the present ULX sample with representative Galactic NS/BH binaries \citep{Makishima2008, Yamada2013, Sasano2015}. 
The comparison is limited to High Mass X-ray Binaries (HMBs), i.e., binaries with massive mass-donating stars, because ULXs are generally considered to be similar systems (e.g., \citealt{Rappaport2005}).
Thus, the Galactic HMXBs exhibit strong emission lines with EW of $20\--300$~eV, and high absorption as $N_{\rm{H}}=1\times10^{21} \-- 7\times 10^{23}$~cm$^{-2}$ which is often highly variable (e.g., two orders of magnitude in Cygnus X-1).
These properties of HMXBs are understood by considering that they are immersed, e.g., in thick stellar winds from the donor stars, and only a small fraction of such circum-source matter is captured by the compact component to drive X-ray emission.
In contrast, the Fe K lines in the ULXs are weaker than $30\--40$~eV in EW, and their $N_{\rm{H}}$ values (after subtracting the Galactic contribution) are all below $8\times10^{21}$~cm$^{-2}$, and variations by $30\%$ at most. 
Thus, in addition to the difference in the absorption lines, figure \ref{fig:n_vs_ew} provides another distinction between ULXs and Galactic HMXBs.
The result suggests that ULXs do not have such massive circum-source materials as seen in the  other class of HMXBs, including stellar winds from the donor stars, of which only a small fraction is captured by the compact component to drive X-ray emission.

A possible explanation to the apparent difference between the ordinary HMXBs and ULXs would be that the matters around ULXs are highly ionized due to their intense X-ray illumination. 
Let us examine this possibility.
Generally, the ionization degree of a gas under illumination with a luminosity $L$ is represented by the ionization parameter as 
\begin{equation}
\label{eq:ionization}
\xi = \frac{L}{nr^{2}}\ ,
\end{equation}
where $n$ and $r$ are the gas density and distance from the X-ray source, respectively. Assuming a gas flow with a nearly-free-fall velocity $v \propto (M/r)^{0.5}$, where $M$ is the object mass, the mass accretion rate can be expressed as
\begin{equation}
\label{eq:windfedmdot}
\dot{m} \propto \pi r^{2} v n \propto \pi r^{1.5}M^{0.5}n.
\end{equation}
Since the luminosity is given as $L=\zeta \dot{m}c^{2}$, where $\zeta$ is the radiation efficiency, equation (\ref{eq:ionization}) can be rewritten, using equation (\ref{eq:windfedmdot}), as
\begin{equation}
\label{eq:ionfinal}
\xi \propto \frac{\zeta n M^{0.5} r^{1.5}}{n r^{2}} \propto \zeta \left( \frac{M}{r} \right)^{0.5}.
\end{equation}
In a wind accretion system, the central object accretes gas within a radius called Bondi radius, where the gravitational pull becomes equivalent to the kinetic energy of the incoming stellar wind. 
This can be written as $R_{\rm{B}} = 2GM/V_{w}^{2}$, where $G$ and $V_{w}$ is the gravitational constant and the velocity of the wind, respectively. 
By substituting $r=R_{\rm{B}}$ in equation~({\ref{eq:ionfinal}}), we obtain the ionization parameter at the distance of Bondi radius as 
\begin{equation}
\label{eq:ion}
\xi \propto \zeta V_{w}~, 
\end{equation}
which is sensitive to neither the luminosity nor the mass of the accreting object. 
Therefore, as long as assuming a wind-fed geometry, we find no reasons to believe that the matters around ULXs are more ionized than those around the ordinary HMXBs. 
The explanation invoking the matter ionization becomes even more difficult if ULXs are under super-critical accretion, because $\zeta$ should be much smaller in such systems (e.g., \citealt{Mineshige2007}).
In short, the matter ionization scenario is not favored.

Another common accretion scheme among binary systems is Roche-Lobe overflow (RLO) accretion, in which the companion star is filling its Roche lobe and matters are accreting through the inner Lagrange point. 
Because the RLO realizes high values of $\dot{m}$, it is favored to explain the high luminosities of ULXs. 
In addition, the absorption and reprocessed spectral features are generally expected to be weaker, since less matters are distributed around the binary system than in wind-fed objects.

LMC X-3 is one of the few good examples of RLO-powered BH binaries in HMXB systems \citep{Soria2001, Orosz2014}. 
It consists of a $7\ M_{\rm{\odot}}$ BH, and a donor star with a mass of $\sim8\ M_{\rm{\odot}}$ \citep{Ozel2010}. 
According to the previous observations, the spectrum of LMC X-3 exhibited low $N_{\rm{H}}$ as $< 10^{20}$~cm$^{-1}$ \citep{Soria2001}, which is comparable to what we have seen in the ULXs. 
On the other hand, the spectrum showed an Fe K$\alpha$ line with an EW ranging between $50\--80$~eV in its Low/Hard state (e.g., \citealt{Nowak2001}, \citealt{Wilms2001}); this is clearly stronger than those in the ULXs. 
Although the example is quite limited, the RLO system in sub-Eddington regimes appears to be somewhat different from ULXs.

How about RLO systems in high-Eddington regimes? 
SMC X-1 is a luminous RLO powered NS HMXB \citep{Savonije1979}. 
Since its luminosity ($3\--5\times 10^{38}$~erg~sec$^{-2}$; e.g., \citealt{Neilsen2004}) is close to or even higher than its $L_{\rm{edd}}$, the source is a good example of a high-$\eta$ RLO system. 
The spectrum of SMC X-1 is relatively featureless, bears weak Fe K$\alpha$ lines with an EW of $20\--30$~eV, and is on average weakly absorbed with $N_{\rm{H}}$ of $2\times 10^{21}$~cm$^{-2}$ (e.g., \citealt{Naik2004}, \citealt{Vrtilek2005}). 
However, $N_{\rm{H}}$ of SMC X-1 is known to vary from $2\times10^{22}$~cm$^{-2}$ to $10^{23}$~cm$^{-2}$ through its super-orbital cycle \citep{Hu2013}, in disagreement with the behavior of ULXs. 
A similar high-$\eta$ NS HMXB is Centaurus X-3, which is also likely to be powered via RLO because of its high luminosity.
Analyzing \textit{Suzaku} data of this pulsar over its full orbital period of 2.1 days, \citet{Naik2011} detected Fe-K line at 6.4~keV (plus higher-ionization lines) throughout the orbital period, and its EW was always $>60$~eV.
Furthermore, even away from the eclipses, the spectra sometimes showed a deep Fe-K absorption edge, which corresponds to $N_{\rm H}$ of several times $10^{23}$ cm$^{-2}$.
These results from SMC X-1 and Centaurus X-3 suggest that high-$\eta$ HMXBs exhibit considerably stronger spectral structures than ULXs, regardless of their accretion scheme (direct wind accretion or RLO).

\subsection{New Possible Accretion Scenario in ULX}
As described so far, the accretion mechanisms commonly seen in HMXB systems may have some difficulties in explaining the basic properties of ULX. This motivated us to look for a new accretion scenario.

\subsubsection{Basic assumptions}
\label{subsub:basic_assumptions}
In general, any explanation of ULXs, either sub-Eddington or super-Eddington, should be able to explain the following conditions.
\begin{itemize}
\item Sufficiently high mass accretion rates should be available, to generate the luminosity of $L_{\rm{X}} = 10^{39-40}$~erg sec$^{-1}$.
\item The accreting compact objects must have a considerable scatter in their masses, at least by an order of magnitude.
\item Except the matter that accretes onto the central objects, the systems should be devoid of excess materials around them. Specially, the low and stable values of $N_{\rm{H}}$ and the upper limits on the Fe-K line EW should be explained.
\end{itemize}
Among these conditions, the second one strongly suggests that ULXs are BHs that are considerably more massive than the ordinary stellar-mass ones.
Although the existence of such BHs used to be unclear, the 10 epic gravitational wave events detected with LIGO and Virgo (section 1), have finally revealed that such BHs as $\sim 50\--80 \ M_{\rm{\odot}}$ do exist, and are probably more abundant than was previously thought. Furthermore, such BHs can grow heavier through repeated mergers.

Although most of the ULX studies have so far implicitly assumed them as mass-exchanging binaries, the third condition above casts doubt on the presence of mass-donating companion stars, particularly massive ones which are suggested by the ULX environment. 
Given these, we revive the idea of \citet{Mii2005}, who interpreted ULXs as single IMBHs, accreting directly via Bondi-Hoyle mechanism from dense parts of the interstellar medium (ISM). 
This scenario agrees with the preferred locations of ULXs, i.e., arms of spiral galaxies.
Furthermore, such a possibility is actually suggested by a recent radio observations by \citet{Oka2016}, that a molecular cloud near Galactic center may be perturbed due to a passage of an isolated BH with a mass of $\sim 10^{5}\ M_{\rm{\odot}}$. Below, let us revisit the scenario of \citet{Mii2005}, considering several properties of ULXs including those found in the present study.

\subsubsection{Luminosity}
Along with the above consideration, let us assume a BH with a mass $M$, entering with a relative velocity $v$ into an ISM cloud which has a mass density $\rho$.
Since the BH captures the ISM within the Bondi radius $R_{\rm{ B}}=2GM/v^2$, the mass accretion rate is described as
\begin{equation}
\dot{M} = \rho \pi R_{\rm{b}}^{2} v = \frac{4\pi G^{2} M^{2} \rho}{v^{3}}~.
\label{eq:mdot}
\end{equation}
Using again the radiation efficiency $\zeta$, equation (\ref{eq:mdot}) can be translated into the luminosity as
\begin{equation}
 L = \zeta \dot{M}c^{2} = \frac{4\pi G^{2} M^{2} c^{2} \zeta \rho}{v^{3}}~.
\label{eq:lbondi}
\end{equation}

Further converting the mass density to the number density
$n=\rho/\mu m_{\rm{p}}$,
where $\mu$ and $m_{\rm{p}}$ are the mean molecular weight and the proton mass, respectively, the luminosity can be expressed as
\begin{equation}
L = 4\times 10^{40}~\zeta_{0.1} ~ {M_{100}}^2 ~ n_{100} ~ v_1^{-3} ~~{\rm erg \ sec^{-1}}.
\label{eq:lbondiscale}
\end{equation}
Here, we assumed $\mu=0.13$, and normalized the relevant quantities as $\zeta_{0.1} \equiv \zeta/0.1$, $M_{100}\equiv M/(100\ M_{\rm{\odot}})$, $n_{100}\equiv n/(100\ {\rm cm^{-3}})$, and $v_1 \equiv v/(1~{\rm km\ s^{-1}})$.
Therefore, if a $100~M_{\rm{\odot}}$ BH is drifting into an ISM cloud of $n\sim 100$ cm$^{-3}$, with a relatively low velocity of $v \sim 1$ km s$^{-1}$ \citep{Nakamura2016}, we expect $L \sim 4\times 10^{40}$~erg~sec$^{-1}$, which is high enough to explain ULXs.

The most important point of this Bondi-Hoyle scenario is the scaling of $L \propto M^{2} v^{-3}$ in equation (\ref{eq:lbondi}).
The mechanism therefore works predominantly for rather massive BHs (as required by the second condition), because of the $M^{2}$ factor, as well as relatively lower values of $v$ which more massive objects would acquire through dynamical friction \citep{Mii2005, Nakamura2016}.

\subsubsection{Absorption and Fe-K lines}
Below, the present scenario is tested against other observational facts.
Let us first assume that a ULX is located at the center of an ISM cloud which has a linear size of
$R_{\rm{ ISM}}=3\times R_3$ pc.
Then, the absorption through the cloud will be
\begin{equation}
N_{\rm{H}} \sim n~ R_{\rm{ISM}} \sim 1 \times 10^{21} \/ n_{100} \/ R_3~~{\rm cm}^{-2}~.
\end{equation}
Assuming $n_{100}\sim 1$ and $R_3 \sim 1$, this agrees with the low and stable absorption observed from the present ULX sample.
For reference, the Bondi radius is $R_{\rm{B}} = 0.9 M_{100}/{v_1}^2$ pc, which is smaller than $R_{\rm{ISM}}$, and the assumed ISM cloud has a mass of
\begin{equation}
M_{\rm{ ISM}} \sim \frac{4\pi}{3} R_{\rm{ ISM}}^3~\rho=300~R_3^3~n_{100}~~M_{\odot}.
\label{eq:ISM_mass}
\end{equation}

Next, we consider the fluorescent Fe-K lines, assuming that the cloud is spherically symmetric, and is illuminated by X-rays from the BH at its center.
In this case, the Fe-K line EW is expected to be approximately proportional to $N_{\rm{H}}$, as indicated by the dashed line in figure \ref{fig:n_vs_ew}.
Thus, when the cloud has $N_{\rm{H}}\sim 10^{21}$~cm$^{-2}$, the Fe-K line EW will be several eV to $\sim 10$~eV, which is fully consistent with the observed upper limits.
This estimate is valid even when the matter within $R_{\rm{ B}}$ is gradually accelerated inwards, as long as the flow can be approximated as spherically symmetric.
The Doppler effect associated with the flow motion would rather reduce the EW of {\em narrow} lines, and would make the scenario even more favorable.

\subsubsection{Activity duration}
Finally, we evaluate two times scales involved in the proposed scenario.
One is the crossing time of the BH through the ISM cloud, expressed as
\begin{equation}
\tau_1 = 2 R_{\rm ISM}/v = 3\times 10^6~R_3/v_1 ~~{\rm yr}~.
\label{eq:tau_1}
\end{equation}
The other is the time scale over which the mass accretion can be sustained.
From equation (\ref{eq:mdot}) and equation (\ref{eq:ISM_mass}), this can be derived as
\begin{equation}
\tau_2 = M_{\rm ISM}/\dot{M} = \frac{(v~R_{\rm ISM})^3}{3G^{2} M^{2}} = 5 \times 10^6~(v_1 R_3)^3/{M_{100}}^2 ~~{\rm yr}~.
\label{eq:tau_2}
\end{equation}
Thus, a single encounter between a massive BH and an ISM clump is estimated to have an activity duration as $\tau_1\sim \tau_2 \sim$ a few million years.
As already argued by \citet{Mii2005}, this time scale is sufficient to power the optical emission-line nebular seen around some ULXs.

As shown so far, the proposed scenario can consistently explain the three conditions raised in section~\ref{subsub:basic_assumptions}.
However, it still remains to be evaluated whether the number of such BHs (with a large uncertainty) and the volume filling factor of rather dense ISM clouds are both high enough to explain the average number of ULXs in each galaxy. 
It is also urgent to observationally examine whether ULXs are indeed imbedded in ISM clumps.
These are left for our future study.

\subsection{Possible accretion regime of ULX}
Although we have so far discussed the possible mass distribution and the accretion mechanism of ULXs, their actual mass accretion regimes, in terms of $\eta$, are still left open to be examined. 
Therefore, let us finally conduct this attempt by referring to the knowledge of BHBs. 
As mentioned in section 4.3, BHBs are known to exhibit distinct spectral states as a function of $\eta$. 
Among several different ways of classification, we adopt that of \citet{Kubota2004}, who identified, in the transient BHB XTE J1550-564, four major spectral states; the Low/Hard state (LHS; $\eta < 0.01$), the High/Soft state (HSS; $\eta \sim 0.1$), the Very High State (VHS; $\eta \sim 0.3$), and the Slim Disk State (SDS; $\eta \sim 1$). 
While \citet{Kubota2004} assigned the SDS literally to the state wherein the accreting matter forms a ``slim disk'' \citep{Abramowicz1988, Watarai2000}, the spectra in this state can be reproduced alternatively, as touched in section 3.2.1, in terms of Comptonization, wherein the disk photons get Comptonized by a cool and optically-thick ($\tau \sim 10$) corona (e.g., \citealt{Miyawaki2009,  Middleton2011}). 
In the present discussion, we adapt this interpretation of the SDS.

Since the spectra of ULX, all significantly Comptonized, were successfully characterized by $Q$ and $yF$, we compare these parameters with those of the individual states of BHB, to seek for any common characteristics between the two classes of objects. 
Hereafter, we exclude the HSS and LHS of BHBs from our discussion, because the former shows no clear sign of thermal Comptonization, and the latter is known to exhibit $T_{\rm{e}}\sim 100$~keV (e.g., \citealt{Makishima2008}) which much exceeds those of ULXs ($T_{\rm{e}}\sim2$~keV).
In addition, the values of $\eta$ in the LHS would be too low to explain ULXs.
\begin{figure}
	\begin{center}
	\includegraphics[width=\columnwidth]{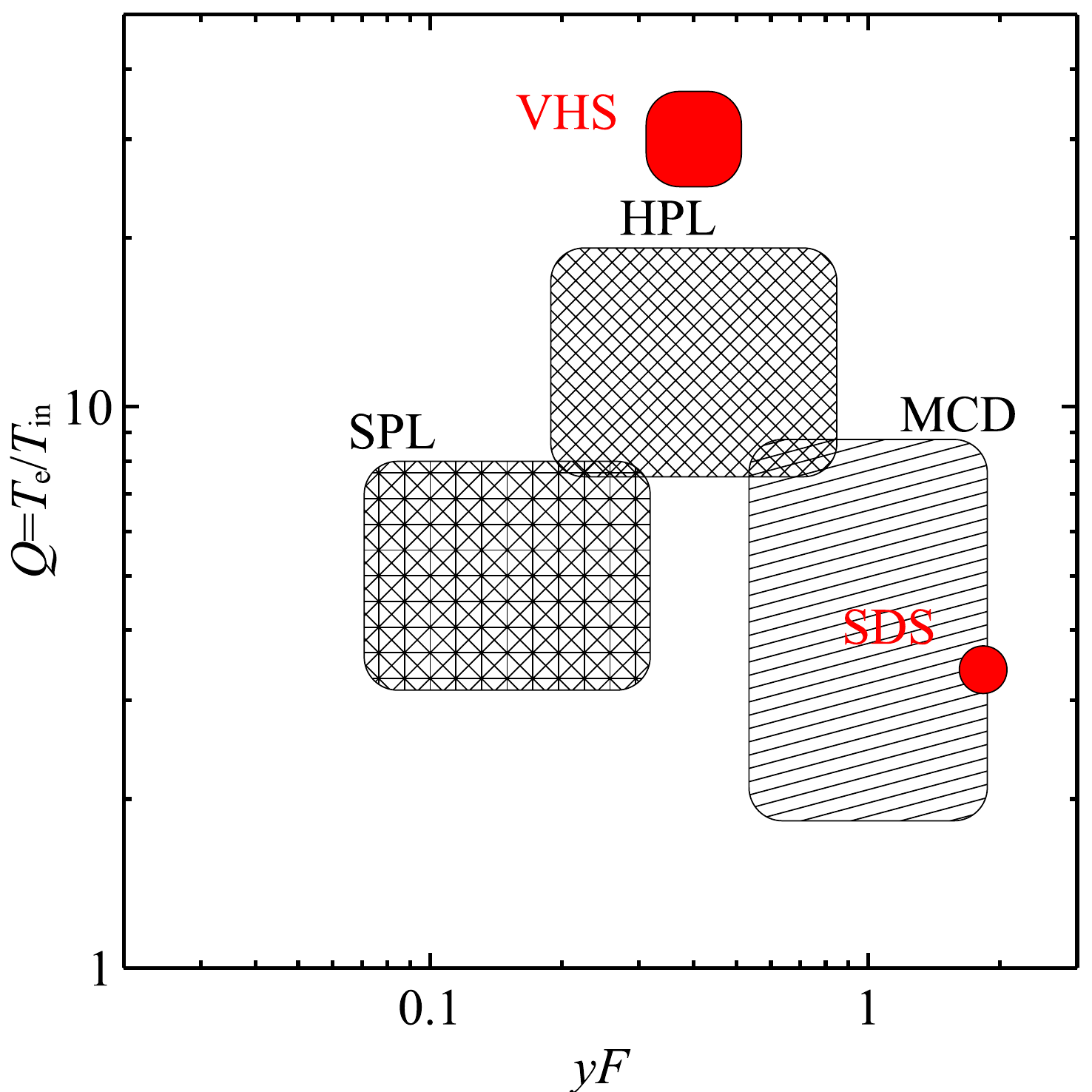}
	\caption{Typical locations of the spectral states of ULXs (black) and BHBs (red), on the $yF$ vs. $Q$ plane. See text for details.}
	\label{fig:ulx_bhb}
	\end{center}
\end{figure}
Figure \ref{fig:ulx_bhb} presents, on the $(yF,\ Q)$ plane, rough locations of these spectral states of ULXs and BHBs. As detailed below, the parameters of BHBs were calculated or estimated utilizing results from several previous studies, including \citet{Done2006}, \citet{Tamura2012}, and \citet{Hori2014} for the VHS, as well as \citet{Kubota2004} for the SDS. Using this figure, let us attempt to find possible correspondence between BHBs and ULXs in terms of their spectral states.

Since few reports have so far been made on Comptonization modeling of BHB spectra in the SDS, we simulated a typical SDS spectrum of XTE J1550-564 obtained by \citet{Kubota2004}, and fitted it with the present MCD+THC model, to obtain $(yF,\ Q)\sim(2,\ 3)$.
Therefore, on the $(yF,\ Q)$ plane, the SDS spectra of BHBs occupy a very similar region to the MCD state of ULX. 
This agreement is very understandable, because these two states are both characterized by strongly convex spectra. 
From these similarities, the MCD state of ULXs can be securely regarded as a spectral state corresponding to the SDS of BHBs, when the latter is considered to harbor Comptonized disks rather than slim disks.
Then, like the SDS, the MCD state is implied to have $\eta \sim 1$.
This implication is fully consistent with our interpretation of ULXs as BHs that are more massive than the ordinary stellar-mass BHs, because the maximum luminosity observed in the present ULX sample, $\sim 10^{40}$ erg sec$^{-1}$, is close to $L_{\rm{edd}}$ of $\sim 100~M_{\rm{\odot}}$ BHs.

The Compton continuum of VHS is typically soft as $\Gamma=2.3\--2.6$ and rolls over at $\sim 40$~keV, which yield $T_{\rm{e}}\sim20$~keV and $y=0.5\--0.7$. 
At the same time, the VHS spectra show a strong soft excess component peaking at $\sim2$~keV, which is considered as the directly visible accretion-disk emission, with $T_{\rm{in}}\sim0.5$~keV and $F=0.6\--0.8$. 
As a result, the VHS spectra are located typically at $(yF,\ Q)\sim(0.4,\ 30)$. 
Thus, the $yF$ values of the VHS fall on a range similar to those of the SPL and HPL of ULXs, although the values of $Q$ are rather different. 
Since they have similar values of $yF$ in figure \ref{fig:ulx_bhb}, and similar continuum slopes ($\Gamma \sim2.3\--2.6$ of the VHS vs. $\Gamma \sim 2.4$ of the SPL state), we may assign the SPL state of ULXs to the VHS of BHBs.
In addition, the SPL luminosities, which are $5\--10$ times lower than that of the MCD state (figure \ref{fig:l_vs_qandyf} D), is in a broad agreement with the values of $\eta\sim0.3$ found in the VHS.

As described above, the SPL state of ULXs exhibited several times lower $Q$ than those in the VHS of BHBs, and the difference could be attributed to the BH mass difference.
In fact, when evaluated at the same $\eta$ and the same distance scaled by the Schwarzschild radius, i.e., at the same intensity to be observed, the coronal electron density $n_{\rm{e}}$ should be inversely proportional to the BH mass if the accretion flow is radially self similar.
Therefore, towards more massive systems, the Compton cooling of coronal electrons (with a rate $\propto n_{\rm{e}}$) would dominate their Coulomb heating by hot ions ($\propto n_{\rm{e}}^2$), resulting in a lower value of $Q=T_{\rm{e}}/T_{\rm{in}}$.
As supporting evidence, the ``soft excess'' structures seen in active galactic nuclei have recently been described successfully as thermal Comptonization in coronae with even lower temperature as $T_{\rm{e}} = 0.2\--0.5$~keV (e.g., \citealt{Noda2011, Done2012, Jin2016}), although these massive BHs should have much lower $\eta$.

Our final issue is how to interpret the HPL state, wherein the Comptonization is further enhanced from the SPL state \citep{Kobayashi2017}.
From the above two assignments, it is natural to consider that the HPL state is a new spectral state (possibly with $0.4 < \eta< 1$), which would fall, in BHBs, between the VHS and SDS.
Although it is unclear at present why the corresponding state is not clearly observed from BHBs, an interesting observation was reported by \citet{Hu2018}; they found that an extragalactic X-ray transient, called NGC 7793 P9, evolved from a presumably HSS of BHBs, into the SPL state, and further into the HPL and MCD states of ULXs.
It is possible that the spectral state sequence of accreting BHs is not completely determined by $\eta$, but depends to some extent on the BH mass itself.

\section*{Acknowledgements}
The authors would like to thanks all the member of the \textit{Suzaku}, \textit{XMM-Newton}, and \textit{NuSTAR} teams for their devotion to instrumental developments, calibration, and spacecraft operation. This present research has been financed by JSPS KAKENHI Grant Numberg JP 18H05873.


\bibliographystyle{mnras}
\bibliography{bondi} 




\appendix
\section{Modification to the spectral model}
As described in section 3.2.2, several MCD spectra in the present sample, one from IC 342 X-1 on 2004 February 20, and a few from NGC 1313 X-1 (table \ref{tab:fitparam}), should unnaturally low $T_{\rm in}$, and yielded extremely large values of $R_{\rm in}$. In this Appendix, we describe how these cases were treated by applying some modification to our canonical MCD+THC model.

Among the 7 spectra of IC 342 X-1, the one obtained on 2004 February 20 exhibited the anomaly (table \ref{tab:fitparam}). This particular spectrum, at the same time, required a stronger absorption of $N_{\rm{H}}=7.3\times10^{21}$~cm$^{-2}$ than that of 2005 February 10 ($N_{\rm{H}}=2.3\times10^{21}$~cm$^{-2}$), even though these two spectra have nearly identical shapes, as seen in figure \ref{fig:fit_res1} (E). 
We hence considered that the large $R_{\rm in}$ of the 2004 spectrum is an artifact caused by this apparently high absorption, and refitted the spectrum with the column density fixed to the average value ($N_{\rm H}=2.5\times10^{21}$ cm$^{-2}$) obtained from the other data sets. 
As shown in table \ref{tab:fitparam}, the fit slightly worsened, but $T_{\rm{in}}$ became higher ($T_{\rm in}=0.5$~keV) than in the previous fit ($T_{\rm in}=0.16$~keV). The revised value is consistent with that of 2005 ($T_{\rm in} = 0.6$~keV) and that of the MCD states seen in the other sources.

Among the 18 spectra of NGC 1313 X-1, 5 (e.g., 2014 May 27) implied the presence of a large disk. However, unlike the case of IC 342 X-1, these spectra showed similar $N_{\rm{H}}$ to those obtained in the other observations of this object. 
We considered that, these spectra actually harbor an extra soft component at $0.3\--0.8$~keV, and interpret it as emission from the outer part of the accretion disk, where the hot corona is no longer present and hence blackbody photons from this region reach us without upscattered. 
In order to represent this consideration, we modified our modeling by adding an extra {\tt diskbb} model which represents this outer-disk (cool) emission, and let the original {\tt diskbb+nthcomp} component to represent the inner-disk (hotter) region where the photons are Comptonized by the corona. 
The results utilizing this modified model are also shown in table \ref{tab:fitparam} as additions to the original results.  
This modified model have actually improved the fits (table \ref{tab:fitparam}), and brought the inner-disk temperature ($T_{\rm in2}$ in table \ref{tab:fitparam}) to $0.6\--0.9$~keV, which is consistent with those of the other MCD-state spectra in our sample.
Furthermore, the parameters of the outer-disk ($T_{\rm{in1}}$ and $R_{\rm{in1}}$ in table \ref{tab:fitparam}) and those of the inner-disk ($T_{\rm{in2}}$, $R_{\rm{in2}}$) satisfy the temperature vs radius relation expected in the standard accretion physics ($T\propto r^{-0.75}$), where $T$ is the effective temperature of the accretion disk, and $r$ is the radius from the disk center; \citep{Shakura1973}.



\bsp	
\label{lastpage}
\end{document}